\documentclass[%
 reprint,
superscriptaddress,
 amsmath,amssymb,
 aps,
]{revtex4-2}
\usepackage{graphicx}% Include figure files
\usepackage{dcolumn}% Align table columns on decimal point
\usepackage{bm}% bold math
\usepackage{tensor}
\usepackage{svg}
\usepackage{subcaption}
\usepackage{tabularray}
\usepackage{physics}
\usepackage[symbol]{footmisc}
\usepackage[utf8]{inputenc} % usually not needed (loaded by default)
\usepackage{soul}
\usepackage[T1]{fontenc}
\usepackage[%  
    colorlinks=true,
    pdfborder={0 0 0},
    linkcolor=blue
]{hyperref}
\usepackage{url}

\begin{document}
\newcommand{\ruoming}[1]{{\color{blue}{ RP: #1}}}
\newcommand{\xuntao}[1]{{\color{orange}{ XW: #1}}}
\setstcolor{red}
\preprint{APS/123-QED}
\title{Hybrid spin-phonon architecture for scalable solid-state quantum nodes}% Force line breaks with \\
%\thanks{These authors contributed equally to this work.}

\author{Ruoming Peng}
\thanks{These two authors contribute equally to this work.}
\email[Corresponding author: ]{pruoming@gmail.com}
% \homepage{http://www.Second.institution.edu/~Charlie.Author}
\affiliation{3. Physikalisches Institut, ZAQuant, University of Stuttgart, Stuttgart, 70569, Germany}

%  Second institution and/or address\\
%  This line break forced% with \\
% }%
% \affiliation{
%  Third institution, the second for Charlie Author
% }%
\author{Xuntao Wu}
\thanks{These two authors contribute equally to this work.}
\affiliation{Pritzker School of Molecular Engineering, University of Chicago, Chicago, IL 60637, USA}

\author{Yang Wang}
%  \homepage{http://www.Second.institution.edu/~Charlie.Author}
\affiliation{3. Physikalisches Institut, ZAQuant, University of Stuttgart, Stuttgart, 70569, Germany}

\author{Jixing Zhang}
%  \homepage{http://www.Second.institution.edu/~Charlie.Author}
\affiliation{3. Physikalisches Institut, ZAQuant, University of Stuttgart, Stuttgart, 70569, Germany}

\author{Jianpei Geng}
%  \homepage{http://www.Second.institution.edu/~Charlie.Author}
\affiliation{School of Physics, Hefei University of Technology, Hefei, Anhui 230009, China}

\author{Durga Bhaktavatsala Rao Dasari}
%  \homepage{http://www.Second.institution.edu/~Charlie.Author}
\affiliation{3. Physikalisches Institut, ZAQuant, University of Stuttgart, Stuttgart, 70569, Germany}
\affiliation{Max-Planck Institute for Solid-state Research, Stuttgart, 70569, Germany}

\author{Andrew N. Cleland}
\affiliation{Pritzker School of Molecular Engineering, University of Chicago, Chicago, IL 60637, USA}
\affiliation{Center for Molecular Engineering and Material Science Division, Argonne National Laboratory, Lemont, IL 60439, USA}

\author{J\"org Wrachtrup}
%  \homepage{http://www.Second.institution.edu/~Charlie.Author}
\affiliation{3. Physikalisches Institut, ZAQuant, University of Stuttgart, Stuttgart, 70569, Germany}
\affiliation{Max-Planck Institute for Solid-state Research, Stuttgart, 70569, Germany}
%  Authors' institution and/or address\\
%  This line break forced with \textbackslash\textbackslash

%\collaboration{MUSO Collaboration}%\noaffiliation

%\author{Charlie Author}
 %\homepage{http://www.Second.institution.edu/~Charlie.Author}
%\affiliation{
% Second institution and/or address\\
% This line break forced% with \\
%}%
%\affiliation{
% Third institution, the second for Charlie Author
%}%
%\author{Delta Author}
%\affiliation{%
% Authors' institution and/or address\\
% This line break forced with \textbackslash\textbackslash
%}%

%\collaboration{CLEO Collaboration}%\noaffiliation

\date{\today}% It is always \today, today,
             %  but any date may be explicitly specified

\begin{abstract}
Solid-state spin systems hold great promise for quantum information processing and the construction of quantum networks. However, the considerable inhomogeneity of spins in solids poses a significant challenge to the scaling of solid-state quantum systems. A practical protocol to individually control and entangle spins remains elusive. To this end, we propose a hybrid spin-phonon architecture based on spin-embedded SiC optomechanical crystal (OMC) cavities, which integrates photonic and phononic channels allowing for interactions between multiple spins. With a Raman-facilitated process, the OMC cavities support coupling between the spin and the zero-point motion of the OMC cavity mode reaching 0.57 MHz, facilitating phonon preparation and spin Rabi swap processes. Based on this, we develop a spin-phonon interface that achieves a two-qubit controlled-Z gate with a simulated fidelity of $96.80\%$ and efficiently generates highly entangled Dicke states with over $99\%$ fidelity, by engineering the strongly coupled spin-phonon dark state which is robust against loss from excited state relaxation as well as spectral inhomogeneity of the defect centers. This provides a hybrid platform for exploring spin entanglement with potential scalability and full connectivity in addition to an optical link, and offers a pathway to investigate quantum acoustics in solid-state systems.
\end{abstract}

%Use showkeys class option if keyword
                              %display desired
\maketitle
%\tableofcontents
\section{\label{sec:level1}Introduction
%\protect
%\\ %The line
%break was forced \lowercase{via} 
%\textbackslash
%\textbackslash
}

Achieving high-fidelity, high-efficiency quantum state transfer, storage, and entanglement between distant qubits is a challenging prerequisite to realizing hybrid quantum systems \cite{Clerk_Lehnert_Bertet_Petta_Nakamura_2020}. 
Solid-state defects are excellent candidates for long-distance quantum communication since their excited states can be optically accessed for remote interconnect through fiber links \cite{Awschalom2022,Chen2022,knaut2023entanglement}. Moreover, these defects have long-coherence electron and nuclear spins \cite{fuchs2011quantum, Abobeih2018, LeDantec2021, Ruskuc2022, anderson2022five, Gupta2023}, with coherence times even exceeding seconds \cite{Abobeih2018,anderson2022five}, making them ideal for quantum memories. Electron and nuclear spins in solids can interact with each other through dipolar and hyperfine interactions \cite{Childress2006,Bermudez2011,Ruskuc2022}, offering a natural platform for entangling spins for quantum computing and quantum simulations. 

As a result, considerable efforts have been invested in developing solid-state spin defects for quantum applications \cite{Doherty2013,Wang2023PhdThesis}.  For instance, remote photon interference of nitrogen-vacancy (NV) centers in diamond coupled with local nuclear spins have enabled the realization of multi-node quantum networks with impressive memory capabilities \cite{Bradley2022}. Furthermore, combining NV spins with nearby nuclear spin registers offers a promising path towards quantum computing, including time crystals and error-corrected quantum algorithms \cite{randall2021many,Abobeih2022}. Within a NV ensemble, the interaction between NV electron spins and the surrounding nuclear spin bath has been leveraged to simulate thermodynamics, spin diffusion, and critical behavior in condensed matter systems \cite{cai2013large,ho2017critical,choi2017observation,kucsko2018critical,zu2021emergent,davis2023probing}. However, the spatial inhomogeneity of defects within solids poses significant challenges in engineering identical spins for scalable quantum systems \cite{aharonovich2016solid,awschalom2018quantum}. In high-density samples, the lack of realistic individual spin control allows only global control, thereby limiting applications for e.g. quantum simulations \cite{zhou2020quantum,choi2020robust}. Consequently, achieving individual control and entanglement of solid-state spins, and enabling their coupling with more physical degrees of freedom (DOF), are important challenges toward practical quantum applications.

Phonon coupling is ubiquitous among all the quantum systems in solids \cite{OConnell2010,Toyoda2015,golter_coupling_2016,Satzinger2018,Wigger2019,Wollack2022,Ripin2023,Wang2024Cooling}. Phonons can be excited through optomechanical or piezoelectric interactions with high conversion efficiency. They travel with velocities of km/s, which is orders of magnitude slower than electromagnetic waves. As a result, acoustic waves in solids can have frequencies in the GHz range, but with wave packet extents significantly smaller than electromagnetic wave with similar frequencies. Given their ability to interact with different physical DOFs, phonons stand out as important intermediate quantum information carriers that can establish coherent interconnects between distant qubit systems \cite{delsing20192019,bienfait2019phonon,mirhosseini2020superconducting,zivari2022non,zivari2022chip,qiao2023splitting}.

Significant progress has been made in developing quantum acoustics by interfacing phonons with a variety of qubits. For example, phonons, as quanta of the strain fields, can be generated in a piezoelectric substrate by electric modulation through accompanying electrodes, or in an optomechanical system by parametrically optical pumping the phonon sideband. This enables efficient coupling of phonons to a wide range of qubit systems, including defect centers, superconducting qubits, quantum dots, and photons \cite{delsing20192019,whiteley2019spin,mirhosseini2020superconducting}. However, a viable spin-phonon interface at the single-phonon regime for efficient spin entanglement is still missing. Here, we adapt the scheme of spin-phonon interaction to cavity optomechanics, considering a localized phonon mode overlapping with tens of electron spins in a sub-micron region. This hybrid architecture can offer a strong coupling of the individual electron spins to the cavity phonons, and further entangling distant spins through a common phononic bus. By considering feasible parameters for current spin-phonon setups in SiC, here we demonstrate a deterministic controlled-Z gate by engineering the geometric phase of a Raman-facilitated phonon dark state, whose fidelity can be further improved through the implementation of carefully designed optical pulses and refined fabrication techniques. Furthermore, we extend this model to consider larger-scale spin systems, demonstrating the generation of highly entangled multi-spin Dicke states with high fidelities. These states are particularly valuable for applications in quantum metrology and sensing and offer potential applications to quantum error correction \cite{wang2021preparing}.

\section{\label{sec:level1} Spin-phonon interaction}
\begin{figure}[tb]
    \centering
    \includegraphics{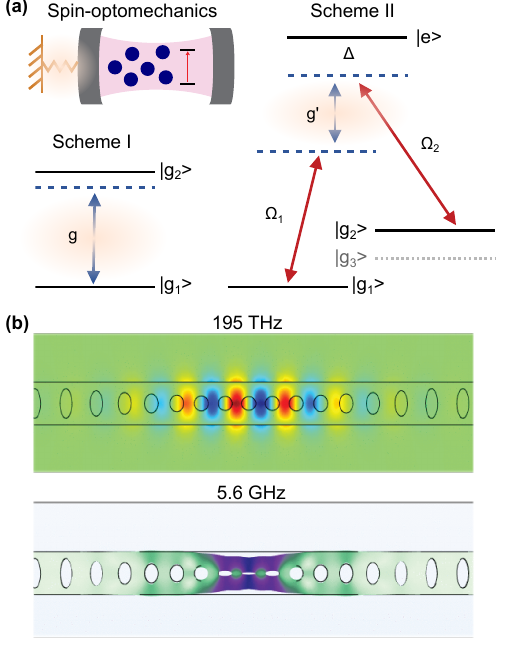}
    \caption{Cavity spin-optomechanics. (a) Illustration of cavity optomechanics with embedded spins. Scheme I denotes the direct spin-phonon coupling and Scheme II shows the enhanced phonon coupling through Raman facilitated process. Here, $g^{\prime}$ denotes the Raman-enhanced spin-phonon coupling strength as opposed to $g$. Two lasers with frequencies $\omega_1, \omega_2$ (not labeled) drive each transitions with Rabi frequencies $\Omega_1$ and $\Omega_2$ respectively, and both drives are offset to the excited state by $\Delta$. The phonon frequency $\omega_m$ as well as the spin transition frequency $\omega_s$ are also not labeled. (b) Finite-element simulation of the optomechanical crystal cavity. The designed SiC cavities host optical cavity resonance at 195 THz within the telecom frequency and phononic resonance at 5.6 GHz.}
    \label{fig:fig1}
\end{figure}

We consider solid-state spins located inside a nanomechanical oscillator, where they can naturally interact with cavity phonons via the strain-induced coupling, as illustrated in scheme \MakeUppercase{\romannumeral 1} of Fig. \ref{fig:fig1}(a). Earlier efforts have shown that the spin states of defects can be controlled by incident phonons when the phonon frequency is near-resonant with the spin transition frequency \cite{whiteley2019spin}. Here we consider dilute spins distributed within the nanomechanical oscillator, in which case no direct intra-spin dipolar interactions are expected.

\subsection{Direct spin-phonon coupling}

When an ensemble of electron spins is placed inside an optomechanical cavity (OMC), the spins can interact with phonons when their frequencies are closely matched. However, the ground-state spin-phonon coupling for most defects is relatively weak, with an estimated zero-point phonon coupling strength below kHz for defects like NV in diamond or Si vacancy in SiC \cite{ovartchaiyapong2014dynamic,meesala2016enhanced,breev2021stress}, while it becomes much stronger for orbital states of group IV defects such as Si or Sn vacancy in diamond \cite{bradac2019quantum,maity2020coherent}. In the latter case, a coupling strength of 40 MHz has been theoretically predicted by mixing spin with orbital states \cite{raniwala2022spin}. While it is possible to reach the strong coupling regime between spin and phonon, the interaction between all spins and phonons will occur simultaneously, making it impractical to control the individual spin dynamics or entangle selected spin pairs. In addition, the orbital states of group IV defects are separated with energy splitting of $\sim50$ GHz for Si vacancy, $\sim830$ GHz for Sn vacancy \cite{bradac2019quantum}, requiring a relatively low working temperature (below 2K) to avoid phonon-induced dephasing.

\subsection{Excited state phonon coupling}
In contrast to the weak phonon coupling observed in spin states, the excited states of defect centers often yield orders of magnitude higher strain-induced coupling strength, arising from the orbital structure. For instance, the excited state phonon coupling in a NV center is characterized to be 1 PHz per unit strain \cite{wang2020coupling}, which is six orders of magnitude higher than the phonon coupling observed for spin states.
The enhancement of spin-phonon coupling in SiC divacancies is even more pronounced, with an excited-state strain modulation of 7 PHz per strain for the PL4 divacancy \cite{falk2014electrically}. Similar to NV centers, divacancies hosted in SiC exhibit exotic spin and optical properties, featuring a spin-1 configuration for their spin states with a five-second electron spin coherence \cite{anderson2022five} and a remarkably bright optical emission rate \cite{anderson2019electrical,miao2019electrically}. Furthermore, their optical linewidth has been improved to 30 MHz at 4 K, approaching the lifetime-limited linewidth via the optimization of annealing and charge depletion processes \cite{anderson2019electrical}. Unlike diamond, which is often challenging to grow and fabricate, SiC is commercially available in the form of low-impurity, single-crystal wafers up to several inches in diameter, and can be easily incorporated into well-established nano-fabrication processes developed for power electronics.

\subsection{Raman-facilitated spin-phonon coupling}
To achieve larger spin-phonon coupling and realize individual control of the spin dynamics in a weak magnetic field environment, we consider the enhanced spin-phonon coupling available through a Raman-facilitated interaction using the excited state of the defect center. 
The Raman scheme was first introduced in trapped-ion systems where the hyperfine states of ions can couple with a common motional mode to achieve all-to-all interactions \cite{steane1997ion}. Here, we follow a similar stimulated Raman process and consider two spin ground states $\ket{g_{1}}$, $\ket{g_{2}}$ connected through the optically excited state $\ket{e}$ of the defect center, forming a $\Lambda$-type system coupled by cavity phonons \cite{golter_coupling_2016} (see scheme \MakeUppercase{\romannumeral 2} of Fig.~\ref{fig:fig1}(a), where $\ket{g_3}$ is an unused third level). The Hamiltonian is then derived with two Rabi drives on the excited state transitions, which is written as
\begin{equation}\label{eq:lambda}
\begin{aligned}
\mathcal{H} = \omega_{m}b^{\dagger}b + \sum\limits_{i}\biggl[
\begin{aligned}[t]
&-\nu_{i1}\ket{g_{i1}}\bra{g_{i1}} - \nu_{i2}\ket{g_{i2}}\bra{g_{i2}}\\
&+ \left( \frac{\Omega_{i1}}{2}e^{-j\omega_{i1}t}\ket{e_{i}}\bra{g_{i1}} + h.c. \right)\\
&+ \left( \frac{\Omega_{i2}}{2}e^{-j\omega_{i2}t}\ket{e_{i}}\bra{g_{i2}} + h.c. \right)\\
&+ g_{i}\left(b^{\dagger} + b\right)\ket{e_{i}}\bra{e_{i}}\biggr],
\end{aligned}
\end{aligned}
\end{equation}
where ${\omega_{i1}}$ and ${\omega_{i2}}$ are the laser drive frequencies with effective field strengths ${\Omega_{i1}}$ and ${\Omega_{i2}}$, and $g_i$ is the excited-state zero-point coupling. Spin transition frequency is then defined as $\omega_{is}=\nu_{i1}-\nu_{i2}$. Here the index $i$ denotes different spins.

As an example, we consider SiC divacancies integrated into an optomechanical crystal (OMC) cavity. The design strategy of the OMC cavity is discussed in the supplementary information \cite{supplement}. Two spin states of the divacancy combine with one optical excited state to form the desired $\Lambda$-type system. As shown in scheme \MakeUppercase{\romannumeral 2} of Fig. \ref{fig:fig1}(a), two drive lasers are configured with a frequency offset $\omega_{1} - \omega_{2}$ close to the spin-phonon detuning $\omega_s-\omega_m$, which are also both detuned by $\Delta=\nu_{i1}-\omega_{i1}=\nu_{i2}-\omega_{i2}$ from the optical transition frequency to avoid actual occupation of the excited state. Thanks to the intrinsic spatial inhomogeneity of the material, excited states of different defect centers can be spectrally distinguished due to crystal dislocation, variations of strain, charge environment, etc. \cite{batalov2009low}, especially if we consider a diluted spin ensemble where individual defect centers exhibit spectral separations larger than both the spin-phonon coupling strength and the laser linewidth. Therefore, by carefully arranging the frequencies of the laser fields, the coupling between phonons and any individual spin can be dynamically controlled. This also applies to single-qubit operations on the spin, where instead we coherently drive the Raman transition within the $\Lambda$ system with a zero frequency offset \cite{golter_optically_2014}. A similar approach has also been applied to controlling the charge state of spins, demonstrating a reversible optical memory beyond the diffraction limit by utilizing their spectral differences \cite{Monge2023}.

As the phonon mode profile is determined by the device structure, we conduct simulations of OMC cavities with varying geometries to investigate the relationship between zero-point coupling $g$ and the phononic mode volume. Surprisingly, even a standard OMC cavity design \cite{chan2012optimized} (see Fig. \ref{fig:fig1}(b)) exhibits a coupling strength of 257 MHz between the excited state of the divacancy and the phonon ground state, which surpasses the expected spin-phonon coupling in a state-of-the-art designed diamond OMC cavity \cite{burek2016diamond}. By implementing an ultra-compact design strategy \cite{raniwala2022spin}, $g$ can be further enhanced, paving the way for even faster operations. It is worth noting that the simulated structure exhibits a co-localization of photonic and phononic modes, providing an additional optomechanical knob to control the phonon population and remotely connect multiple cavities through fiber links.

To estimate the effective spin-phonon coupling $g^{\prime}$ from the large excited state phonon coupling $g$, we apply the Schrieffer-Wolff transformation (see Supplementary Information \cite{supplement} for more details) to simplify the Hamiltonian in Eq.~(\ref{eq:lambda}) to the standard Jaynes–Cummings form. This effective coupling between spin ground states and the phonon is now written as
\begin{equation}\label{eq:eff_cpl}
\mathcal{H}_{\mathrm{int}} =  g\frac{\Omega_{1}\Omega_{2}}{4\left| \Delta\right|\omega_{m}}b^{\dagger}\ket{g_{1}}\bra{g_{2}} + h.c.,
\end{equation}
where $g^{\prime} = g\Omega_{1}\Omega_{2}/4\abs{\Delta}\omega_{m}$ is the effective spin-phonon coupling assisted by the excited state. In this configuration, $g^{\prime}$ arises from the periodic driving of the $\Lambda$ system, which is proportional to both $g$ and the Rabi frequencies of the drives. As $g^{\prime}$ is much smaller than phonon frequency  ${\Omega_{m}}$, contributions from higher phonon occupation states are negligible.

\begin{figure}[tb]
    \centering
    \includegraphics{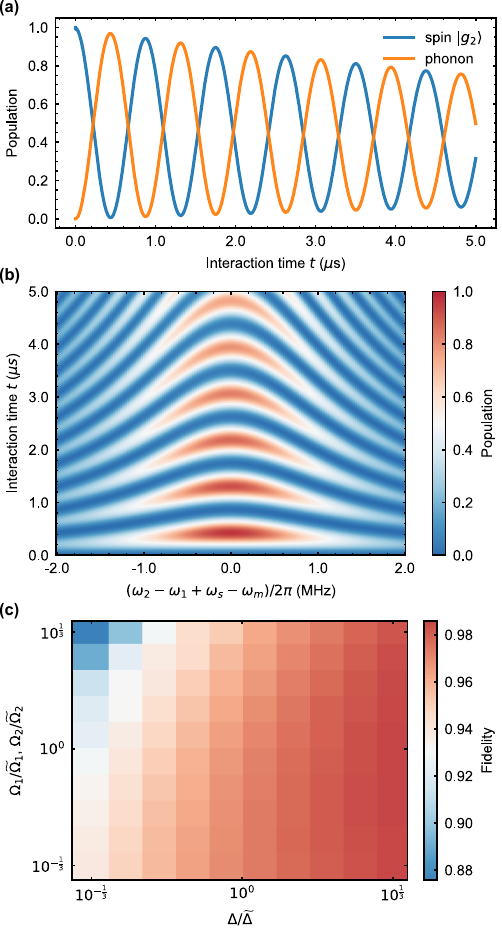}
    \caption{Phonon-facilitated ODRO. (a) Coherent swap between a single spin and the phonon mode. (b) The ``Chevron'' interference pattern, generated by sweeping the frequency offset of two laser drives. (c) Fidelity of single-phonon preparation as a function of $\Delta$, $\Omega_{1}$ and $\Omega_{2}$, where they are scaled relative to $\widetilde{\Delta}/2\pi$=230 MHz, $\widetilde{\Omega}_{1}/2\pi$=500 MHz and $\widetilde{\Omega}_{2}/2\pi$=23 MHz, which are used for the simulations in (a) and (b). \label{fig:fig2}}
\end{figure}

\section{\label{sec:level1} Optically driven spin-phonon interaction
%\protect
%\\ %The line
%break was forced \lowercase{via} 
%\textbackslash
%\textbackslash
}
The main idea of our proposal is that the coupling of spins to phonons can be enhanced by two additional drive lasers, achieving coupling strengths approaching MHz. We consider a coupled spin-phonon system at mK experimental temperatures (assume a microwave frequency of around $5~\mathrm{GHz}$, which gives an effective temperature of about $240~\mathrm{mK}$), ensuring that the thermal phonon population in the OMC cavity is negligible. Even at elevated temperatures, the OMC cavity can be initialized to its phonon ground state via optomechanical interactions~\cite{Chan2011}. Consequently, we consistently start from the phonon ground state for the coupled spin-phonon system, where phonon excitations in the cavity are primarily driven by spin-phonon interactions. We also note that the optical powers required in our scheme are moderate and applied in pulsed sequences only during active gate operations, minimizing steady-state heating. Furthermore, gates can be scheduled to reduce simultaneous activity and manage heat dissipation within a dilution refrigerator.

\subsection{Choices of system parameters}
Here, we take effective Rabi frequencies much smaller than the detuning ($\Omega_{1}/2\pi$=500 MHz, $\Omega_{2}/2\pi$=23 MHz, $\Delta/2\pi$=230 MHz, $\omega_{m}/2\pi$=5.6 GHz). Since the divacancy host bright optical transition, achieving an effective Rabi frequency of 500 MHz requires the laser power of $\sim$200 $\mu$W. Therefore, we can operate the system in dispersive regime with an effective spin-phonon coupling of $g^{\prime}/2\pi = 0.57~\mathrm{MHz}$, according to Eq.~(\ref{eq:eff_cpl}). This leads to the Phonon-facilitated optically driven Rabi oscillation (ODRO) \cite{golter_optically_2014}. Taking into account both the electron spin coherence time and the phononic cavity lifetime exceeding ms, the coupled system resides in the strong coupling regime where $g^{\prime}\gg\Gamma_{s}, \Gamma_{m}$. The intrinsic phonon loss at GHz frequencies in SiC is comparable to that in diamond and lower than in silicon. By carefully optimizing the fabrication process, it is possible to achieve low-loss SiC OMC cavities, reaching performance levels similar to other leading platforms—such as diamond~\cite{huang2025ultralow} and silicon~\cite{maccabe2020nano}. Here, $\Gamma_{s}$ and $\Gamma_{m}$ are the linewidth of electron spin and phononic cavity respectively. In addition to the intrinsic loss of the spin and the phononic cavity, the excited state of the defect center will also introduce extra leakage and decoherence on the order of $\Gamma_e\Omega_1\Omega_2/\abs{\Delta}\omega_m$, where $\Gamma_e$ is the linewidth of the defect's optical transition. Thanks to the dispersive condition, the assumption of the strong coupling regime remains valid. In all subsequent simulations, we model the phonon mode by truncating the Hilbert space to a maximum phonon number of 5. Other parameters for the following simulation are summarized in Table~\ref{tab:ODRO}.

\begin{table*}[tb]
 \caption{\label{tab:ODRO}Simulation parameters (unit: GHz) for resonant ODRO. $\Gamma$ indicates the decoherence part of the system, where the subscript denotes the source of the decoherence ($m$ for the phonon mode, $e$ for the defect's excited state, and $s$ for the defect's spin states), while the superscript represents the type ($1$ for the energy decay and $\phi$ for the pure dephasing).}
 \centering
 \begin{ruledtabular}
 \begin{tabular}{cccccccccc}
   $\omega_{m}/2\pi$&$g/2\pi$&$\Delta/2\pi$&$\Omega_{1}/2\pi$&$\Omega_{2}/2\pi$&$\Gamma_{m}^{1}$&$\Gamma_{e}^{1}$&$\Gamma_{e}^{\phi}$&$\Gamma_{s}^{1}$&$\Gamma_{s}^{\phi}$\\
   \hline
   5.6 & 0.257 & 0.23 & 0.5 & 0.023 & $10^{-6}$ & 0.01~\cite{miao2019electrically} & 0.02~\cite{miao2019electrically} & $10^{-9}$~\cite{anderson2022five} & $10^{-6}$~\cite{anderson2022five}\\
 \end{tabular}
 \end{ruledtabular}
\end{table*}

\subsection{State transfer between spin and phonon}
For the theoretical analysis, we consider the coupling of the divacancy spin states $\ket{0}$ ($\ket{g_1}$) and $\ket{+1}$ ($\ket{g_2}$) through the excited state, and set the $\ket{-1}$ ($\ket{g_3}$) state decoupled from the Raman driving protocols. When the frequency offset of driving lasers matches the spin-phonon detuning, i.e. $\omega_1-\omega_2=\omega_s-\omega_m$, a coherent vacuum Rabi oscillation between spin state $\ket{g_2}$ and the phonon occurs as illustrated in Fig. \ref{fig:fig2}(a). Using parameters in Table~\ref{tab:ODRO}, the hybrid spin-phonon system already exhibits a large cooperativity $C=g^{\prime2}/\Gamma_s\Gamma_m\approx3.2\times10^5$ and the overall fidelity for one-phonon state preparation reaches $96.82\%$, comparable to other qubit-phonon interaction systems.

Furthermore, we plot a ``Chevron'' interference pattern by sweeping the frequency offset of two laser drives, as shown in Fig. \ref{fig:fig2}(b). This pattern demonstrates the control of the spin-phonon dynamics at the single-phonon level in the OMC cavity. 
According to Eq.~(\ref{eq:eff_cpl}), the spin-phonon coupling is proportional to the Rabi frequencies of the laser drives. Intuitively, larger laser power can result in a higher gate fidelity owing to the enhancement in the coupling strength. However, the intrinsic excited state decoherence of the defect center is also magnified by the strong laser drives and thereby adds additional decoherence to the hybrid system. On the other hand, if we naively decrease the laser drive power or increase the excited state detuning $\Delta$, the evolution will become too slow so that other decoherence sources will dominate. As shown in Fig. \ref{fig:fig2}(c), the fidelity of single-phonon preparation is simulated as a function of $\Delta$ and the Rabi frequencies where the overall fidelity is saturated to be $98.59\%$.

\begin{figure}[htbp]
    \centering
    \includegraphics{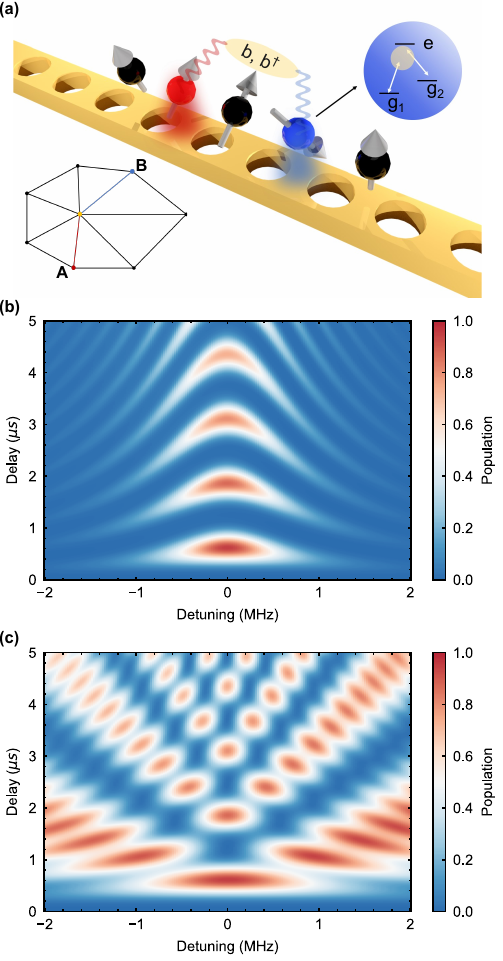}
    \caption{Phonon-facilitated ODRO between two spins. Prepare $\ket{A,B} = \ket{g_2,g_1}$ and measure the population in $\ket{A,B} = \ket{g_1,g_2}$ as a function of spin detuning and delay. (a) Diagram showing the two spin-phonon coupling schemes. The active spins $A$ and $B$ are highlighted in red and blue which are connected to the common phononic channel by Raman facilitated process, while other spins denoted by the black sphere are inactive and remain ``dark'' to the phonon. (b) ``Chevron'' interference pattern as we tune the two spin frequencies in the opposite direction w.r.t. the phonon mode. (c) Population swapping between the two spins as we keep them aligned but vary the common detuning to the phonon mode. Interference between distant qubits is revealed by the detuning of the qubit frequency.  \label{fig:fig3}}
\end{figure}

In addition to the high spin-phonon entanglement fidelity, the large cooperativity also leads to the capability of high-precision single-shot readout for the spin state.
% Since the strongly coupled system is within the regime of the low cavity dissipation, we achieve that the coupling strength is larger than effective spin decoherence and much larger than the phonon loss: $g^{\prime}>\Gamma_{op}\gg\Gamma_{m}$. 
When the frequency offset of the laser drives is far detuned from the spin-phonon detuning, i.e. $\abs{\omega_1-\omega_2}\gg\abs{\omega_s-\omega_m}$, the coupled spin-phonon system resides in the dispersive regime, where the spin state of the defect induces a $2\chi$ frequency shift of the phonon resonance with $\chi=g^{\prime2}/\left( \omega_1-\omega_2-\omega_s+\omega_m \right)$. In such a scenario, by probing the cavity phonon response, we are able to distinguish between the defect spin states as a consequence of the spin-phonon coupling. Additionally, such a system provides the required optomechanical interaction where single-shot readout is also attainable through optomechanical-induced transparency (OMIT), as discussed in the recent report \cite{koppenhofer2023single}. Here, we have engineered the optical resonance into the telecom regime to enable potential optical communication via the optomechanical channel. It is important to note that the optical resonance of the cavity can also be tailored closer to the divacancy transition energy at $\sim$1100 nm by slightly reducing the size of the OMC unit cells. In this way, we can enhance the optical interaction and achieve faster operation speeds in the Raman scheme. Thanks to the Raman facilitated coupling scheme, we are able to achieve both individual control and readout in the spin ensemble since excited-state transitions of the spins are spectrally distinguished due to the unavoidable inhomogeneity in solids.

\subsection{Phonon bus for spin interaction}
Given that a single spin strongly couples to the cavity phonon, the phonon mode can then be utilized as a bus to entangle distant spins.
As illustrated in Fig. \ref{fig:fig3}(a), the input laser can be tuned by an optical frequency shifter (OFS) to match the frequency detuning of each spin, as well as the individual control of the laser amplitude. Multi-channel microwave tones independently mix with the common laser input such that the phase and amplitude of each driving laser beam can be separately controlled, providing an operating bandwidth of more than 100 GHz using the state-of-the-art OFS \cite{hu2021chip}. This wide bandwidth can effectively compensate for the considerable spatial inhomogeneity in a SiC OMC cavity, enabling selective entanglement between arbitrary spin pairs. Such many-to-many connectivity is beneficial in designing efficient quantum error correction protocols \cite{Bravyi2024, Xu2024}.

In our simulation, we consider two distant spins labeled $A$ and $B$ (with negligible direct dipolar interaction) to be independently controlled by two sets of laser beams. The two spins are first initialized to the states $\ket{g_2}$ and $\ket{g_1}$, respectively. Then the corresponding laser beams are configured as mentioned before to connect each spin with the zero-point fluctuation of the cavity phonon mode. 

By applying parameters in Tabel~\ref{tab:ODRO}, now to both spins, a coherent population swap between spin $A$ and $B$ can be achieved. As shown in Fig. \ref{fig:fig3}(b), we simulate the interaction between spin $A$ and $B$ when they are detuned in opposite directions with respect to the phonon resonance. A similar ``Chevron'' type oscillation is observed, with a state transfer fidelity of $94.92\%$ when both spins are on resonance, corresponding to an iSWAP gate operating point. Together with single-qubit gates, arbitrary quantum operations can be implemented in the coupled spin-phonon system, enabling universal quantum computing. When both spins are detuned in the same direction, another interference pattern of the state population is obtained (see Fig. \ref{fig:fig3}(c)). This indicates a transition from on-resonance to virtual phonon interaction in the OMC cavity.

\begin{figure*}[htbp]
    \centering
    \includegraphics{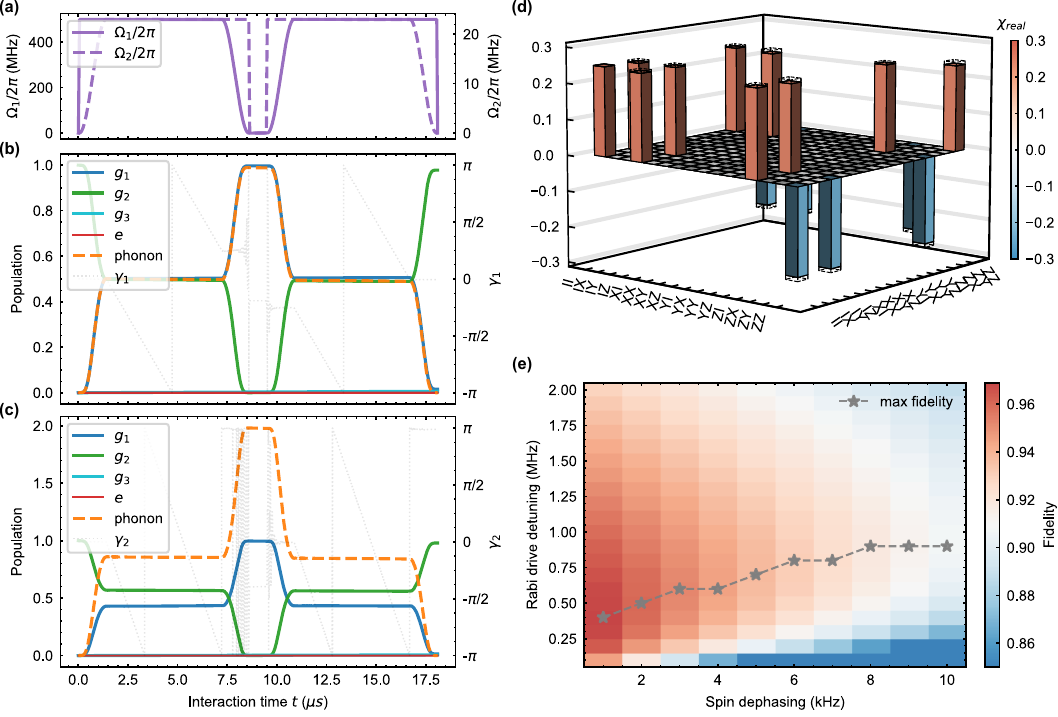}
    \caption{STIRAP process for the two-qubit Controlled-Z (CZ) gate implementation. (a) STIRAP pulse sequence, where the majority of population transfer occurs during the rising and dropping stages enclosed by the ``square'' pulse shape. Here we only plot the absolute values of both driving amplitudes, see Supplementary Information \cite{supplement} for more details. (b) Population and phase evolution when only one spin is coupled to the STIRAP process. Phonon is pumped to its first excited state and then transferred back again, which preserves the $\ket{g_2g_3}$ and $\ket{g_3g_2}$ states. (c) Population and phase evolution when two spins are coupled to the STIRAP process simultaneously. Phonon is pumped to its second excited state and then transferred back again, yielding a $\gamma_2-2\gamma_1=\pi$ phase difference for the $\ket{g_2g_2}$ input state compared to the single spin scenario. (d) Full quantum process tomography for the CZ gate demonstrated by the real part of the $\chi$ matrix, featuring a gate fidelity of 96.80\%. (e) Demonstration of the feasibility of the gate protocol with some non-ideal system parameters (e.g. the spin dephasing here), over 90\% gate fidelity, is still achievable with spin coherence time down to $100~\mu s$. \label{fig:fig4}}
\end{figure*}

\section{\label{sec:level1}Adiabatic evolution via phonon dark-state
%\protect
%\\ %The line
%break was forced \lowercase{via} 
%\textbackslash
%\textbackslash
}

We have shown that strong coupling between the spin and the phonon mode is achievable while the leakage to the excited state can be suppressed by increasing the laser detuning. While another significant decoherence source of the excited state comes from the spectral diffusion of the optical transition. Throughout the ODRO process, such spectral diffusion perturbs the effective coupling strength $g^{\prime}$, and degrades the operation fidelities. An efficient way to mitigate such decoherence is to evolve the system through an adiabatic process so that the system is always trapped in a so-called dark state, which is more robust against the frequency shift induced by the spectral diffusion \cite{golter_optically_2014}. More details of comparison between ODRO and STIRAP in terms of their robustnesses against spectral diffusion is provided in Supplementary Information~\cite{supplement}.

This adiabatic protocol is termed stimulated Raman adiabatic passage (STIRAP) \cite{STIRAP_RMP,vitanov_stimulated_2017} and has been proposed as effective in constructing geometric phase gate \cite{STIRAP_PRA2007} and realizing multi-ion entanglement \cite{STIRAP_PRL2008}. Therefore, it allows us not only to reduce population leakage to the unwanted states but also to achieve precise control of the phase for certain states through the adiabatic process, manifesting the phase-related gate implementation and state generation. Notably, STIRAP has been successfully deployed for single-qubit gate operations in NV centers, achieving a fidelity of up to 93\% \cite{Zhou2017accelerated}.
% Furthermore, there is potential for additional enhancement, potentially achieving performance levels comparable to those achieved through microwave control.
Here, our focus is primarily on the STIRAP process involving the phonon as a quantum field, with the Hamiltonian written as
\begin{equation}\label{eq:dark_state:general}
\begin{aligned}
\mathcal{H} = &- \left( \frac{\Omega_{1}g}{2\omega_m}b\sum\limits_{i}\ket{e_{i}}\bra{g_{i1}} + h.c. \right)\\
&+ \left( \frac{\Omega_{2}}{2}\sum\limits_{i}\ket{e_{i}}\bra{g_{i2}} + h.c. \right).
\end{aligned}
\end{equation}
Here we assume that the laser detuning ($\Delta$ in Fig.~\ref{fig:fig1}(a)) is zero to maintain an ideal adiabatic pathway, and the driving amplitudes $\Omega_1, \Omega_2$ are identical across all the defects, see Supplementary Information \cite{supplement} for more details. This holds promises for realizing high-fidelity qubit gates as well as genuine entanglement within a spin ensemble.

\subsection{CZ gate}

As has been mentioned before, we consider a $\Lambda$ system (see Fig.~\ref{fig:fig1}(a)) formed with the defect's ground states $\ket{0}$, $\ket{+1}$ and excited state $\ket{e}$. A phonon-assisted drive occurs on the $\ket{0}\leftrightarrow\ket{e}$ transition. Once again, we denote $\ket{0}$ ($\ket{+1}$) as $\ket{g_{1}}$ ($\ket{g_{2}}$), with $\ket{+1}$ defined as the qubit one state $\ket{1_q}$. There is another ground state $\ket{-1}$ ($\ket{g_{3}}$) decoupled from both laser drives so that it can be treated as the qubit zero state $\ket{0_q}$ as no phase would be accumulated in this state. Phonon states will be denoted using numbers equivalent to its Fock level. Therefore, the basis state of the spin-phonon system can be written as $\ket{ns_1s_2\cdots s_i\cdots s_N}$, where $n$ is the phonon number and $s_i\in\{ g_1,g_2,g_3,e \}$ are the states for a total of $N$ defects.

In the case of a single spin, the dark state in the one-excitation subspace is defined as
% \begin{equation}
%     \ket{D} = \frac{\Omega_{2}\ket{1g_{1}} + \Omega_{R}\ket{0g_{2}}}{\sqrt{\abs{\Omega_{2}}^{2} + \abs{\Omega_{R}}^{2}}}
% \end{equation}
\begin{equation}
    \ket{D_1} = \Omega_{2}\ket{1g_{1}} + \Omega_{R}\ket{0g_{2}}
\end{equation}
% \begin{equation}
%     \ket{D} = \frac{\frac{\Omega_{2}}{2\Omega_{R}}\ket{2g_{1}g_{1}} + \frac{\Omega_{R}}{\sqrt{2}\Omega_{2}}\ket{0g_{2}g_{2}} + \frac{1}{\sqrt{2}}\left( \ket{1g_{1}g_{2}} + \ket{1g_{2}g_{1}} \right)}{\sqrt{\abs{\frac{\Omega_{2}}{2\Omega_{R}}}^{2} + \abs{\frac{\Omega_{R}}{\sqrt{2}\Omega_{2}}}^{2} + 1}}
% \end{equation}
up to a normalization factor, where $\Omega_{R} = \Omega_{1}g/\omega_{m}$ is the effective Rabi frequency for the phonon sideband transition. If we initialize the system in the state $\ket{0g_{2}}$ and then follow the first half of the pulse sequence illustrated in Fig. \ref{fig:fig4}(a), the population would be adiabatically transferred to the phonon mode, which yields the state $\ket{1g_{1}}$ (see Fig. \ref{fig:fig4}(b) at around 9 $\mu s$). To suppress the excited state leakage and maintain the adiabatic passage in the dark state manifolds, the rising rate of the Rabi drives ($1/t_{\mathrm{rise}}$, where $t_{\mathrm{rise}}$ is the rising time) should be small compared to their amplitudes through the state transfer periods. In addition to the population swap, the dark state will also pick up a non-vanishing geometric phase $\gamma_{1} = -\int \cos^{2}{\theta} d\phi$ (see Supplementary Information \cite{supplement}).

\begin{table}[tb]
 \caption{\label{tab:STIRAP}Simulation parameters (unit: GHz, except $t_{\mathrm{rise}}$) for STIRAP-based CZ gate. Note that instead of using constant values for $\Omega_{1}$ and $\Omega_{2}$, we set the detuning $\Delta=0$ and incorporate a rising time $t_{\text{rise}}$ (time spent to reach the target Rabi frequency) for the adiabatic process. We've eliminated the pure dephasing of the excited state since the STIRAP process is robust against such noises~\cite{supplement}. Other decoherence parameters are the same as in Table.~\ref{tab:ODRO}.}
 \centering
 \begin{ruledtabular}
 \begin{tabular}{cccc}
   $\omega_{m}/2\pi$&$g/2\pi$&$\Delta/2\pi$&$t_{\mathrm{rise}}$\\
   \hline
   5.6 & 0.257 & 0 & 1.35~$\mu s$\\
 \end{tabular}
 \end{ruledtabular}
\end{table}

In another scenario, where we collectively couple two spins to the common phonon mode, the dark state will evolve in the two-excitation subspace. Adjusting the respective laser drive parameters can make the two ions indistinguishable in terms of their phonon interaction, which means we can just consider the populations in the symmetric subspace of the coupled spin-phonon system. In this case, the spin-phonon dark state is calculated as
\begin{equation}
\begin{aligned}
    \ket{D_2} = {}&\frac{\Omega_{2}}{2\Omega_{R}}\ket{2g_{1}g_{1}} + \frac{\Omega_{R}}{\sqrt{2}\Omega_{2}}\ket{0g_{2}g_{2}}\\
    &+ \frac{1}{\sqrt{2}}\left( \ket{1g_{1}g_{2}} + \ket{1g_{2}g_{1}} \right).
\end{aligned}
\end{equation}
Likewise, if we evolve from $\ket{0g_{2}g_{2}}$, the overall phase accumulated during the adiabatic process is $\gamma_{2} = -\int 2\left( 1 - \sin^{4}{\theta} \right) / \left( 2 - \cos^{4}{\theta} \right) d\phi$ (see Supplementary Information \cite{supplement}). Note that the phase difference
\begin{equation}
    \delta \gamma = \gamma_{2} - 2\gamma_{1} = \int \frac{2\cos^{4}{\theta}\sin^{2}{\theta}}{2 - \cos^{4}{\theta}}d\phi
\end{equation}
is non-trivial during the overlap region of $\Omega_{R}$ and $\Omega_{2}$, so that in principle a two-qubit phase gate could be implemented by careful design of the pulse parameters. We design a two-qubit controlled-Z (CZ) gate by having a time-reversed symmetric pulse shape, as illustrated in Fig. \ref{fig:fig4}(a). In this case, the phonon population is transferred back to the spin subsystem at the end of the sequence, which results in a pure phase difference with respect to the initial state. We show the population and phase evolution of one (Fig. \ref{fig:fig4}(b)) and two (Fig. \ref{fig:fig4}(c)) spins interacting with the phonon mode in an adiabatic passage, which amounts to evolving the two qubit states $\ket{1_q0_q}/\ket{0_q1_q}$ and $\ket{1_q1_q}$ respectively. 

We find that by precise control of the adiabatic process, it is possible that a $\pi$ phase is only accumulated when the qubits are in their $\ket{1_q1_q}$ state (see the gray dotted line in Fig. \ref{fig:fig4}). In Fig. \ref{fig:fig4}(d), we show the full process tomography of the CZ gate, with fidelity of $96.80\%$ using reasonable decoherence for both the phonon mode and the divacancy centers, demonstrating the feasibility for precise many-body quantum control in the divacancy spin-phonon system using the STIRAP protocol. Higher fidelities are achievable by incorporating quantum control toolkits to optimize the pulse shapes in Fig. \ref{fig:fig4}(a) and also for improving cavity performance and spin coherence, as shown in the simulation in Fig. \ref{fig:fig4}(e). 

In addition, we have tested the robustness of these processes when a parameter of the system is degraded, such as by including spin dephasing, see Fig. \ref{fig:fig4}(e). Interestingly, we can manipulate the detuning between the two Rabi drives so that the total gate time can be decreased in case of a larger spin dephasing, recognizing there is an upper limit as a large Rabi drive detuning will likely violate the adiabatic condition, which could cause unfavorable excited state populations. Overall, we can still maintain around 90\% gate fidelity even as the spin and/or phonon coherence times drop to around 100 $\mu s$. Therefore, together with the single-qubit gates readily achievable by Raman-type optical drives \cite{Zhou2017accelerated}, we now have protocols for implementing arbitrary multi-qubit quantum gates in coupled spin-phonon systems.

It is worth noting that, while both the Raman (ODRO) and STIRAP schemes can realize two-qubit gates in our system, they differ in their sensitivity to experimental imperfections. The Raman scheme is more susceptible to spectral diffusion of the spin-optical transition, which can lead to reduced fidelity. In contrast, STIRAP offers greater robustness in this regard, as the final state is governed by the pulse trajectory rather than precise interaction timing. However, STIRAP imposes more stringent requirements on the pulse shaping and relative phase control of the optical drives, making experimental implementation more demanding.

\subsection{Multi-spin entangled state}

\begin{figure}[tb]
    \centering
    \includegraphics{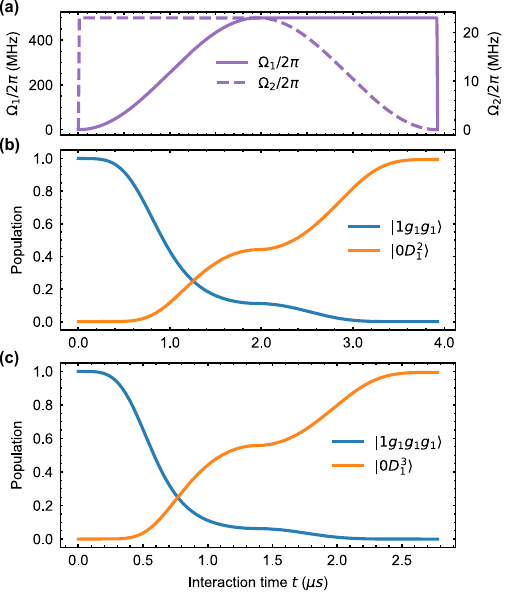}
    \caption{STIRAP process for the generation of multi-spin one-excitation Dicke states $D_{1}^{N}$, where $N$ is the number of spins. (a) Pulse sequence for the adiabatic process, similar to the first half of Fig.~\ref{fig:fig4}(a), but without the center plateau. (b) Generation of $D_{1}^{2} = \left( \ket{g_2g_1} + \ket{g_1g_2} \right)/\sqrt{2}$ in 3927 ns, with a fidelity $\mathcal{F} = \mathrm{tr}\left( \rho_{\mathrm{ideal}}\rho \right) = 99.35\%$. (c) Generation of $D_{1}^{3} = \left( \ket{g_2g_1g_1} + \ket{g_1g_2g_1} + \ket{g_1g_1g_2} \right)/\sqrt{3}$ in 2777 ns, with a fidelity $\mathcal{F} = \mathrm{tr}\left( \rho_{\mathrm{ideal}}\rho \right) = 99.36\%$. System parameters are the same as the implementation of the CZ gate. \label{fig:fig5}}
\end{figure}

In the previous section, we demonstrate the interaction scheme where spins can be entangled through a common phonon mode, allowing for selective connection and operation of arbitrary spins.  The established individual control and all-to-all connectivity of spins are important in the development of more efficient quantum error correction protocols \cite{wang2023faulttolerant,Bluvstein2023,Bravyi2024,Xu2024}. Although the resulting two-qubit gate fidelity of $96.80\%$ is still insufficient to practically benefit from fault-tolerance and error correction \cite{devitt2013quantum,Terhal2015RMP}, this can be improved by optimization of the device fabrication and spin properties. More importantly, it demonstrates an efficient way to engineer the Hamiltonian of the spins in the cavity even with very realistic parameter settings. This engineering capability is important for realizing a practical quantum advantage using near-term noisy devices by implementing applications such as quantum simulation \cite{Daley2022,Cai2013} and quantum machine learning \cite{Martinez2021,Bravo2022}.
Here, we further demonstrate the usefulness of the established control method by developing an efficient scheme for the preparation of highly entangled spin states, which is typically a prerequisite for many quantum applications. 

In particular, we consider the preparation of Dicke states, which are robust against various noise sources \cite{zhang2014quantum} and therefore hold great potential for many applications, including quantum sensing \cite{Saleem2024PRA}, and computing \cite{Ouyang2014}. Some previous preparation schemes rely on global control of the spins by a superconducting transmon qubit \cite{Marcos2010}. However, the coupling between the transmon and the spins is typically inhomogeneous, making thick state preparation experimentally infeasible \cite{pezze2018quantum,hakoshima2020efficient,wang2021preparing}. On the other hand, gate-based preparation schemes would require a large number of gates for even a few tens of spins, which would ultimately limit the efficiency of the preparation process, and therefore the fidelity of the entangled state \cite{Mukherjee2020,bartschi2019deterministic}.

The Raman driving protocols discussed earlier can facilitate the independent control of each spin, therefore enabling the interaction between multiple spins and the phonon mode simultaneously to construct highly entangled spin states. Given that each spin can be accessed and manipulated independently, it is possible to adjust the effective frequency offset of the spins as well as their coupling strengths to the phonon mode to be identical, which allows us to generate Dicke states with an arbitrary spin number. In addition, STIRAP protocol can also be readily integrated into such process by careful design of the driving amplitudes. As a result, the dynamics of the system is fully characterized by Eq.~(\ref{eq:dark_state:general}). Similarly, since all the spins are identical, it is justified to restrict the calculation within the symmetric bases. More details of the model are included in supplementary Information \cite{supplement}.

The simulation results are shown in Fig.~\ref{fig:fig5}, where the OMC cavity is initialized with a single phonon occupation, while all the spins are initialized in $\ket{g_1}$. As driving lasers are on, all the spins interact with a phonon simultaneously through a collective dark state. We simulate the system with 2 and 3 spins inside the cavity, both demonstrating fidelities of Dicke states above $99\%$. As we increase the spin number in our cavity, the preparation of the Dicke state becomes faster given that the effective coupling strength scales with the number of spins with a factor of $\sqrt{N}$ owing to the superradiance effect~\cite{Dicke1954}. Following our protocols, a multi-spin Dicke state can be easily prepared, where the maximum spin number is only limited by the number of spectral distinguished spins we can find inside the OMC cavity.

\section{\label{sec:level1}Conclusion and outlook
%\protect
%\\ %The line
%break was forced \lowercase{via} 
%\textbackslash
%\textbackslash
}
In summary, we have proposed a hybrid spin-optomechanical architecture that hosts an enormous strain-induced excited state modulation for divacancy centers in SiC, from which we have demonstrated strong spin-phonon coupling assisted by a Raman-facilitated process, with one-phonon preparation fidelity up to $96.82\%$. We have further shown that the ability to perform individual spin-phonon interactions can facilitate Rabi swaps between different spins, which still presents a formidable challenge in solid-state spin systems.

Moreover, we have incorporated the STIRAP protocol into our design framework, resulting in a two-qubit CZ gate featuring a fidelity up to $96.80\%$ in the divacancy spin system with current state-of-the-art system parameters, leveraging the involvement of a single phonon mode --- a novel approach that, to the best of our knowledge, hasn't been previously explored in spin ensembles lacking direct interaction. The ability of individual control and coupling also facilitates the preparation of highly coherent quantum states, such as the multi-spin Dicke states with over $99\%$ fidelities, which are crucial for promoting quantum applications in sensing and simulations. Current fidelities are constrained by fabrication imperfections and material inhomogeneity. Improvements in cavity design, for instance, through more compact phononic crystals \cite{raniwala2022spin}; and in spin coherence, via isotopic purification or optimized implantation depth, offer realistic paths toward achieving higher fidelities. Additionally, the use of quantum optimal control techniques can further suppress leakage and off-resonant transitions, enhancing overall gate performance.

Importantly, our proposed scheme is capable of serving as an intermediate quantum node for diverse physical platforms, owing to the nature of the phonon that it can be coupled to nearly any physical DOFs. For instance, given this profound phononic coupling enhanced by the Raman facilitated process, the network of spins can be accessed by other types of qubits, such as superconducting qubits, where the transduction of microwave to optical photon can serve as another interesting candidate to explore in our scheme \cite{mirhosseini2020superconducting,shandilya2021optomechanical}. In addition, following earlier reports on coupled superconducting-bulk acoustic wave resonators \cite{chu2017quantum}, higher phonon number states can also be prepared similarly using our scheme, which is a crucial step to realize an error-protected bosonic mode \cite{bild2023schrodinger}. Last but not least, the strong flexibility of our approach allows for the realization of distributed quantum systems utilizing solid-state spins, which can be achieved by connecting different OMC cavities, each hosting spectrally distinguished spins, through an expandable phononic or optomechanical network integrated with optical control and readout. Our study thus provides new perspectives on using solid-state spin systems as novel quantum information processing resources and is applicable to many other physical platforms given versatile quantum control paradigms and interfaces.

\begin{acknowledgments}
R.P. and X.W. contributed equally to this work.\newline 

We acknowledge the financial support by the Baden-Württemberg Foundation via project SPOC, the German Research Foundation (DFG) via project FOR 2724, the Carl Zeiss Foundation via the Center for Integrated Quantum Science and Technology, and the Max Planck School of Photonics. Y.W. and D.B.D. were supported by the German Federal Ministry of Education and Research (BMBF) through the project QECHQS with Grant No. 16KIS1590K. X.W. and A.N.C. were supported by the U.S. Army Research Office and the Laboratory for Physical Sciences (ARO grant W911NF2310077), by the U.S. Air Force Office of Scientific Research (AFOSR grant FA9550-20-1-0270), and in part by the U.S. Department of Energy Office of Science National Quantum Information Science Research Centers.

\end{acknowledgments}

\begin{quote}

\end{quote}

\clearpage
\onecolumngrid
\begin{center}
\textbf{\large Supplementary Information: Hybrid spin-phonon architecture for scalable solid-state quantum nodes}
\end{center}
\setcounter{section}{0}
\renewcommand{\thesection}{S\arabic{section}}
\setcounter{figure}{0} % Resets the figure counter to 0
\renewcommand{\thefigure}{S\arabic{figure}} % Changes the figure numbering to S1, S2, etc.
\renewcommand{\figurename}{Fig.} % Ensures the caption starts with "Fig." instead of "Figure"
\setcounter{table}{0}
\renewcommand{\thetable}{S\arabic{table}}%
\setcounter{equation}{0} % Reset equation numbering
\renewcommand{\theequation}{S\arabic{equation}} % Prefix with S. and Arabic number

\section{Design of SiC optomechanical crystal cavity}

The design of the optomechanical crystal (OMC) cavities builds upon the previously established approach developed in Silicon. Each cavity consists of a one-dimensional nanobeam composed of periodic ``mirror'' cells, which gradually transition through an adiabatic, Gaussian-like taper into a central ``defect'' cell, as depicted in the main text. The nominal unit cell of the SiC mirror region, along with its design parameters (a, t, w, hx, hy)=(580, 250, 705, 220, 488) nm, is illustrated in Figure S1. The corresponding parameters for the central defect cell are (a, t, w, hx, hy)=(436, 250, 705, 265, 227) nm. To achieve co-localization of both photonic and phononic modes within the cavity, we optimized the mirror cell unit parameters, enabling the formation of a large photonic and phononic band gap, as demonstrated in Figure S1.
\begin{figure}[htbp]
    \centering
    \includegraphics{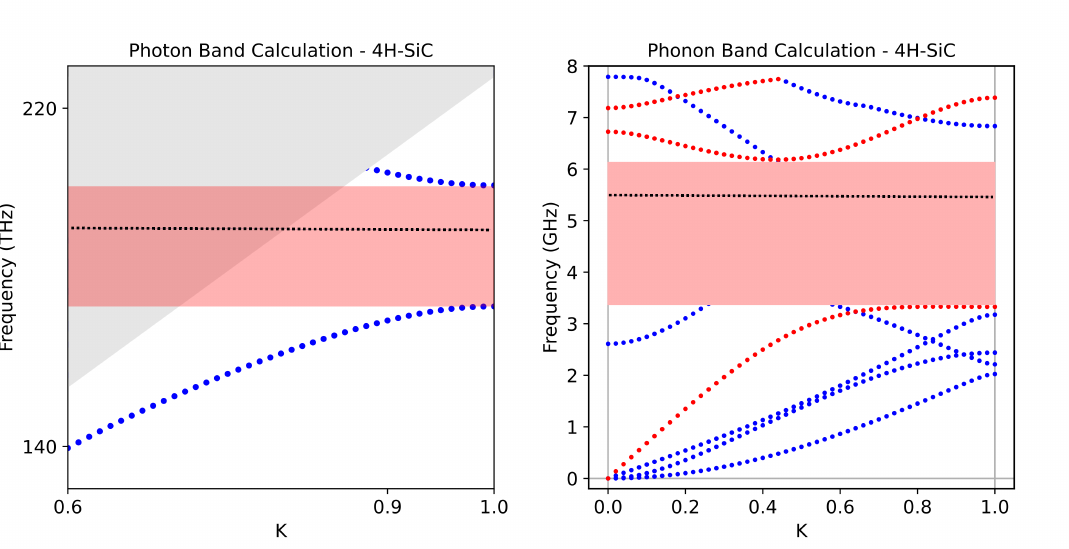}
    \caption{Figure (left)  presents the photonic band dispersion of the SiC OMC cavity, which features a photonic bandgap exceeding 30 THz. The gray-shaded region above the band structure indicates a continuum of radiation and leaky modes, while the pink-shaded region marks the quasi-photonic bandgap. The cavity resonance at 195 THz is highlighted by a black dashed line. Figure (right) depicts the phononic band structure of the OMC cavity, showing a phononic bandgap of 3 GHz. The y- and z-symmetric mode is marked by a red dot, with other vector symmetries represented by blue dots.\label{fig:band_dispersion}}
\end{figure}

\section{Estimate of coupling rates in the SiC OMC cavity}
Here, we quantify the key interactions that are critical for controlling our hybrid system. Within the optomechanical cavity, the spin can couple with cavity phonons. In the main text, we have analyzed the interaction strengths for both the spin ground state and excited state. Based on previous studies of divacancies in SiC, the energy shift for the spin ground state is on the order of a few GHz per strain unit, whereas the energy shift for the spin excited state is significantly stronger, reaching 7 PHz per strain unit for PL4 divacancies. Using finite element simulations, we estimate that the zero-point fluctuation of the OMC cavity induces a strain of approximately $3.7\times 10^{-8}$. Therefore, the exited state phonon coupling is 257 MHz and the ground state spin phonon coupling is about 35 Hz.

The engineered OMC cavity also exhibits strong optomechanical interactions, driven by both the moving boundary effect and the photoelastic effect. This structure offers a powerful means of manipulating the phonon state within the system, enabling techniques such as optomechanical cooling to achieve the phonon ground state. Recent proposals also suggest that hybrid spin-optomechanical devices could be utilized for efficient spin-state readout.

\begin{table*}[tb]
 \caption{\label{tab:OMCcoup_rate} Coupling rates for optomechanical control. Here, $\omega_{p}/2\pi$ and $\omega_{m}/2\pi$ represent the photonic and phononic frequencies of the cavity mode, respectively. The terms $g_{MB}/2\pi$ and $g_{PE}/2\pi$ denote the optomechanical coupling rates associated with the moving boundary effect and the photoelastic effect. The parameter $Q_r$ reflects the radiative loss of the designed OMC cavity.
}
 \centering
 \begin{ruledtabular}
 \begin{tabular}{cccccccccc}
   $\omega_{p}/2\pi$&$\omega_{m}/2\pi$&$g_{MB}/2\pi$&$g_{PE}/2\pi$&$Q_r$\\
   \hline
   195 THz & 5.6 GHz & 358 KHz & 84 KHz & 6.7 M\
 \end{tabular}
 \end{ruledtabular}
\end{table*}

\section{Defect-phonon coupling model}
Here, we only show the derivation process of a collection of lambda systems coupled to a common phonon mode, for systems that have a larger number of states as a manifold the method is similar. For lambda systems, we have a generic coupled Hamiltonian in the lab frame
\begin{equation}
\begin{aligned}
\mathcal{H} = {}& \omega_{m}b^{\dagger}b + \sum\limits_{i}\biggl[
\begin{aligned}[t]
&-\nu_{i1}\ket{g_{i1}}\bra{g_{i1}} - \nu_{i2}\ket{g_{i2}}\bra{g_{i2}} + g_{i}\left(b^{\dagger} + b\right)\ket{e_{i}}\bra{e_{i}}\\
&+ \left( \frac{\Omega_{i1}}{2}e^{-j\omega_{i1}t}\ket{e_{i}}\bra{g_{i1}} + h.c. \right) + \left( \frac{\Omega_{i2}}{2}e^{-j\omega_{i2}t}\ket{ei}\bra{g_{i2}} + h.c. \right) \Biggr].
\end{aligned}
\end{aligned}
\end{equation}
Then, consider the phonon coupling term $g_{i}\left( b^{\dagger} + b \right)\ket{e_{i}}\bra{e_{i}}$ as a perturbation and use the same generator $\mathcal{S} = \sum_{i}g_{i}/\omega_{m}\left( b^{\dagger} - b \right)\ket{e_{i}}\bra{e_{i}}$ as in \cite{golter_coupling_2016}, we can perform Schrieffer-Wolff transformation \cite{bravyi_schriefferwolff_2011} so as to get a series of effective coupling to the optical transitions of the divacancy centers
\begin{equation}
\begin{aligned}
\mathcal{H^{\prime}} ={}& \mathcal{UHU^{\dagger}} = e^{\mathcal{S}}\mathcal{H}e^{-\mathcal{S}}\\
={}& \mathcal{H} + \left[\mathcal{S}, \mathcal{H}\right] + \frac{1}{2}\left[ \mathcal{S}, \left[ \mathcal{S}, \mathcal{H} \right] \right] + \cdots\\
={}& \omega_{m}b^{\dagger}b + \sum\limits_{i}
\begin{aligned}[t]
&\left[ -\nu_{i1}\ket{g_{i1}}\bra{g_{i1}} - \nu_{i2}\ket{g_{i2}}\bra{g_{i2}} - \frac{g_{i}^{2}}{\omega_{m}}\ket{e_{i}}\bra{e_{i}} \right.\\
&\ +\left( \frac{\Omega_{i1}}{2}e^{\frac{g_{i}}{\omega_{m}}\left( b^{\dagger} - b \right) - j\omega_{i1}t}\ket{e_{i}}\bra{g_{i1}} + h.c. \right)\\
&\ + \left.\left( \frac{\Omega_{i2}}{2}e^{\frac{g_{i}}{\omega_{m}}\left( b^{\dagger} - b \right) - j\omega_{i2}t}\ket{e_{i}}\bra{g_{i2}} + h.c. \right) \right]
\end{aligned}\\
{}&-2\sum\limits_{i<j}\frac{g_{i}g_{j}}{\omega_{m}}\ket{e_{i}}\bra{e_{i}}\otimes\ket{e_{j}}\bra{e_{j}}.
\end{aligned}
\end{equation}
Here without loss of generality, we let the $\ket{g_{i1}}\leftrightarrow \ket{e_{i}}$ transition couple to the first phonon sideband with beam-splitter-type interaction, while the $\ket{g_{i2}}\leftrightarrow \ket{e_{i}}$ transition couple to the carrier band of the Raman process. In this way, if we further set detunings for each individual transition, we can obtain the linearized Hamiltonian in the rotating frame
\begin{equation}\label{eq:defect-phonon}
\begin{aligned}
\mathcal{H}^{\prime}_{\mathrm{int}} = {}&\sum\limits_{i}\left[ -\Delta_{i1}\ket{g_{i1}}\bra{g_{i1}} - \Delta_{i2}\ket{g_{i2}}\bra{g_{i2}} - \left( \frac{\Omega_{iR}}{2}b\ket{e_{i}}\bra{g_{i1}} + h.c. \right) + \left( \frac{\Omega_{i2}}{2}\ket{e_{i}}\bra{g_{i2}} + h.c. \right) \right]\\
{}&-2\sum\limits_{i<j}\frac{g_{i}g_{j}}{\omega_{m}}\ket{e_{i}}\bra{e_{i}}\otimes\ket{e_{j}}\bra{e_{j}},
\end{aligned}
\end{equation}
where $\Omega_{iR} = \Omega_{i1}g_{i}/\omega_{m}$ is the reduced phonon-assisted Rabi frequency. Note that here we have already assumed a beam-splitter interaction between the phonon mode and the optical transition ($b\otimes \ket{e_{i}}\bra{g_{i1}}$) so that the two-mode squeezing interaction ($b^{\dagger}\otimes \ket{e_{i}}\bra{g_{i1}}$) would be suppressed since it is detuned by twice the phonon frequency.
\section{Optically-driven Rabi Oscillation}

% \begin{table}
%  \caption{\label{tab:table1}Simulation parameters (unit: GHz) for Fig. 2 and 3 in the main text. $\Gamma$ indicates the decoherence part of the system, where the subscript denotes the source of the decoherence ($m$ for the phonon mode, $e$ for the defect's excited state, and $s$ for the defect's spin states), while the superscript represents the type ($1$ for the energy decay and $\phi$ for the pure dephasing).}
%  \centering
%  \begin{ruledtabular}
%  \begin{tabular}{cccccccccc}
%    $\omega_{m}/2\pi$&$g/2\pi$&$\Delta/2\pi$&$\Omega_{1}/2\pi$&$\Omega_{2}/2\pi$&$\Gamma_{m}^{1}$&$\Gamma_{e}^{1}$&$\Gamma_{e}^{\phi}$&$\Gamma_{s}^{1}$&$\Gamma_{s}^{\phi}$\\
%    \hline
%    5.6 & 0.257 & 0.23 & 0.5 & 0.023 & $10^{-6}$ & 0.01 & 0.02 & $10^{-9}$ & $10^{-6}$\\
%  \end{tabular}
%  \end{ruledtabular}
% \end{table}

Here we will start with Eq.~(\ref{eq:defect-phonon}) to derive an effective spin-phonon coupling Hamiltonian in the large detuning scenario. In fact, when the detunings are sufficiently large, the excited state population can be neglected, and so as the effective excited state coupling term (the second line in Eq.~(\ref{eq:defect-phonon})), which reduces our Hamiltonian to
\begin{equation}\label{eq:odro}
\begin{aligned}
\mathcal{H} = {}& \mathcal{H}_{\mathrm{diag}} + \mathcal{H}_{\mathrm{offdiag}}\\
{}&\sum\limits_{i}\left( -\Delta_{i1}\ket{g_{i1}}\bra{g_{i1}} - \Delta_{i2}\ket{g_{i2}}\bra{g_{i2}}\right) + \sum\limits_{i}\left[ - \left( \frac{\Omega_{iR}}{2}b\ket{e_{i}}\bra{g_{i1}} + h.c. \right) + \left( \frac{\Omega_{i2}}{2}\ket{e_{i}}\bra{g_{i2}} + h.c. \right) \right].
\end{aligned}
\end{equation}
We can define the generator $\mathcal{S}$ for the Schrieffer-Wolff transformation $\mathcal{H}^{\prime} = e^{\mathcal{S}}\mathcal{H}e^{-\mathcal{S}}$ as
\begin{equation}\label{eq:odro-generator}
\mathcal{S} = \sum\limits_{i}\left( -\frac{\Omega_{iR}}{2\Delta_{i1}}b\ket{e_{i}}\bra{g_{i1}} + \frac{\Omega_{i2}}{2\Delta_{i2}}\ket{e_{i}}\bra{g_{i2}} + h.c. \right)
\end{equation}
such that $\mathcal{H}_{\mathrm{offdiag}} + [\mathcal{S}, \mathcal{H}_{\mathrm{diag}}] = 0$ and the transformed Hamiltonian $\mathcal{H}^{\prime}$ will be block-diagonalized to the second order
\begin{equation}\label{eq:odro-SW}
\begin{aligned}
\mathcal{H}^{\prime} \simeq {}& \mathcal{H}_{\mathrm{diag}} + \frac{1}{2}\left[ \mathcal{S}, \mathcal{H}_{\mathrm{offdiag}} \right]\\
= {}&\sum\limits_{i}
\begin{aligned}[t]
\Biggl[& -\left( \Delta_{i1} + \frac{\abs{\Omega_{iR}}^{2}}{4\Delta_{i1}}b^{\dagger}b \right)\ket{g_{i1}}\bra{g_{i1}} - \left( \Delta_{i2} + \frac{\abs{\Omega_{i2}}^{2}}{4\Delta_{i2}} \right)\ket{g_{i2}}\bra{g_{i2}}\\
{}&+ \frac{\Omega_{iR}^{*}\Omega_{i2}}{8}\left( \frac{1}{\Delta_{i1}} + \frac{1}{\Delta_{i2}} \right)b^{\dagger}\ket{g_{i1}}\bra{g_{i2}} + \frac{\Omega_{iR}\Omega_{i2}^{*}}{8}\left( \frac{1}{\Delta_{i1}} + \frac{1}{\Delta_{i2}} \right)b\ket{g_{i2}}\bra{g_{i1}} \Biggr].
\end{aligned}
\end{aligned}
\end{equation}
This resembles the form of the Dicke model where multiple two-level-systems (TLSs) coherently couple to a common quantum field mode. In the case of a single defect, this reduces to the standard Jaynes-Cummings model where a coherent population transfer can occur between the phonon mode and the spin.

\section{Dark-state calculation}
Let us reconsider Eq.~(\ref{eq:odro}) where
\begin{equation}\label{eq:dark_state}
\mathcal{H} = \sum\limits_{i}\left( -\Delta_{i1}\ket{g_{i1}}\bra{g_{i1}} - \Delta_{i2}\ket{g_{i2}}\bra{g_{i2}}\right) + \sum\limits_{i}\left[ - \left( \frac{\Omega_{iR}}{2}b\ket{e_{i}}\bra{g_{i1}} + h.c. \right) + \left( \frac{\Omega_{i2}}{2}\ket{e_{i}}\bra{g_{i2}} + h.c. \right) \right].
\end{equation}
Now, instead of focusing on the large detuning regime, we adiabatically alter the amplitudes and phases of the Rabi drives such that the system is always trapped in its instantaneous ground state by a relatively large spectral gap. To formulate this dynamics and the way to engineer the Controlled Phase gate, we first consider the scenario where a single spin couples to the phonon mode, followed by the case where two identical spins are involved at the same time.
\subsection{Single spin}
When a single spin is coupled to the phonon mode, we have
\begin{equation}\label{eq:dark_state:single:H}
\mathcal{H} = -\Delta_{1}\ket{g_{1}}\bra{g_{1}} - \Delta_{2}\ket{g_{2}}\bra{g_{2}} - \left( \frac{\Omega_{R}}{2}b\ket{e}\bra{g_{1}} + h.c. \right) + \left( \frac{\Omega_{2}}{2}\ket{e}\bra{g_{2}} + h.c. \right).
\end{equation}
If we assume a zero-temperature environment for the phonon mode (which means the phonon can only gain population through coherent transfer with the spin), it is justified to consider in the subspace spanned by $\left\{ \ket{1g_{1}}, \ket{0g_{2}}, \ket{0e} \right\}$. In this case, Eq.~(\ref{eq:dark_state:single:H}) can be expressed in the matrix form (letting $\Delta_{1} = \Delta_{2} = 0$, which means a resonance condition)
\begin{equation}\label{eq:dark_state:single:mat}
\mathcal{H} = \frac{1}{2}
\begin{pmatrix}
    0 & 0 & -\Omega_{R}^{*}\\
    0 & 0 & \Omega_{2}^{*}\\
    -\Omega_{R} & \Omega_{2} & 0
\end{pmatrix},
\end{equation}
from which we can derive a dark state \cite{golter_coupling_2016}
\begin{equation}\label{eq:dark_state:single:D1}
\begin{aligned}
\ket{D_{1}} &= \frac{\Omega_{2}\ket{1g_{1}} + \Omega_{R}\ket{0g_{2}}}{\sqrt{\abs{\Omega_{2}}^{2} + \abs{\Omega_{R}}^{2}}}\\
&= e^{j\phi}\cos{\theta}\ket{1g_{1}} + \sin{\theta}\ket{0g_{2}},
\end{aligned}
\end{equation}
where $\theta$ and $\phi$ are defined such that \cite{STIRAP_PRA2007}
\begin{equation}\label{eq:dark_state:single:theta_phi}
\begin{aligned}
\Omega_{R}(t) &= \sin{\theta(t)}\sqrt{\abs{\Omega_{R}(t)}^{2} + \abs{\Omega_{2}(t)}^{2}},\\
\Omega_{2}(t) &= \cos{\theta(t)}\sqrt{\abs{\Omega_{R}(t)}^{2} + \abs{\Omega_{2}(t)}^{2}}e^{j\phi(t)}.
\end{aligned}
\end{equation}
Since the dark state has zero energy, the overall phase accumulated is the geometric phase
\begin{equation}\label{eq:dark_state:single:geo_phase}
\gamma_{1}^{g} = j\int\ev**{\left(\bm{e}_{\theta}\frac{\partial}{\partial\theta} + \bm{e}_{\phi}\frac{\partial}{\partial\phi}\right)}{D_{1}}\cdot \left( \bm{e}_{\theta}\partial\theta + \bm{e}_{\phi}\partial\phi \right) = j\int\ev**{\frac{\partial}{\partial\phi}}{D_{1}}\partial\phi = -\int\cos^{2}{\theta}\partial\phi.
\end{equation}
\subsection{Two identical spins}
When two identical spins are coupled to the phonon mode simultaneously, we have
\begin{equation}\label{eq:dark_state:two:H}
\mathcal{H} = \sum\limits_{i=1,2}\left[-\Delta_{1}\ket{g_{i1}}\bra{g_{i1}} - \Delta_{2}\ket{g_{i2}}\bra{g_{i2}} - \left( \frac{\Omega_{R}}{2}b\ket{e_{i}}\bra{g_{i1}} + h.c. \right) + \left( \frac{\Omega_{2}}{2}\ket{e_{i}}\bra{g_{i2}} + h.c. \right)\right].
\end{equation}
Again assuming a zero-temperature environment, we will consider in the subspace spanned by $$\left\{ \ket{2g_{1}g_{1}}, \ket{0g_{2}g_{2}}, \ket{0ee}, \left( \ket{1g_{1}g_{2}} + \ket{1g_{2}g_{1}} \right)/\sqrt{2}, \left( \ket{1g_{1}e} + \ket{1eg_{1}} \right)/\sqrt{2}, \left( \ket{0g_{2}e} + \ket{0eg_{2}} \right)/\sqrt{2} \right\}.$$
Here we also take into account the fact that the two spins are identical, which leads to the bases symmetric upon exchange of spin indices. Therefore, the matrix form of Eq.~(\ref{eq:dark_state:two:H}) reads
\begin{equation}\label{eq:dark_state:two:mat}
\mathcal{H} = 
\begin{pmatrix}
    0 & 0 & 0 & 0 & -\Omega_{R}^{*} & 0\\
    0 & 0 & 0 & 0 & 0 & \Omega_{2}^{*}/\sqrt{2}\\
    0 & 0 & 0 & 0 & -\Omega_{R}/\sqrt{2} & \Omega_{2}/\sqrt{2}\\
    0 & 0 & 0 & 0 & \Omega_{2}^{*}/2 & -\Omega_{R}^{*}/2\\
    -\Omega_{R} & 0 & -\Omega_{R}^{*}/\sqrt{2} & \Omega_{2}/2 & 0 & 0\\
    0 & \Omega_{2}/\sqrt{2} & \Omega_{2}^{*}/\sqrt{2} & -\Omega_{R}/2 & 0 & 0
\end{pmatrix},
\end{equation}
from which we can also solve for the dark state
\begin{equation}\label{eq:dark_state:two:D2}
\begin{aligned}
\ket{D_{2}} &= \frac{\frac{\Omega_{2}}{2\Omega_{R}}\ket{2g_{1}g_{1}} + \frac{\Omega_{R}}{\sqrt{2}\Omega_{2}}\ket{0g_{2}g_{2}} + \frac{1}{\sqrt{2}}\left( \ket{1g_{1}g_{2}} + \ket{1g_{2}g_{1}} \right)}{\sqrt{\abs{\frac{\Omega_{2}}{2\Omega_{R}}}^{2} + \abs{\frac{\Omega_{R}}{\sqrt{2}\Omega_{2}}}^{2} + 1}}\\
&= \frac{\cos^{2}{\theta}}{\sqrt{2-\cos^{4}{\theta}}}e^{j\phi}\ket{2g_{1}g_{1}} + \frac{\sqrt{2}\sin^{2}{\theta}}{\sqrt{2-\cos^{4}{\theta}}}e^{-j\phi}\ket{0g_{2}g_{2}} + \frac{2\sin{\theta}\cos{\theta}}{\sqrt{2-\cos^{4}{\theta}}}\frac{1}{\sqrt{2}}\left( \ket{1g_{1}g_{2}} + \ket{1g_{2}g_{1}} \right),
\end{aligned}
\end{equation}
where $\theta$ and $\phi$ satisfy the same conditions as in Eq.~(\ref{eq:dark_state:single:theta_phi}). Note that there exists another dark state
\begin{equation}\label{eq:dark_state:two:D2tilde}
\begin{aligned}
\ket{\widetilde{D}_{2}} = {}&\frac{1}{\sqrt{\left( 
3 + 2\cos^{2}{\theta} - 3\cos^{4}{\theta} \right)\left( 
2 - \cos^{4}{\theta} \right)}}\biggl[ -\sqrt{2}\sin^{2}{\theta}\ket{2g_{1}g_{1}} + e^{-j\cdot2\phi}\left( 
2\cos^{4}{\theta} - 3\cos^{2}{\theta} \right)\ket{0g_{2}g_{2}}\\
&{}+ e^{-j\phi}\sqrt{2}\sin{\theta}\cos{\theta}\left( 
1 + \sin^{2}{\theta} \right)\frac{1}{\sqrt{2}}\left( 
\ket{1g_{1}g_{2}} + \ket{1g_{2}g_{1}} \right) + \left( 
2 - \cos^{4}{\theta} \right)\ket{0ee}\biggr]
\end{aligned}
\end{equation}
for Eq.~(\ref{eq:dark_state:two:mat}), but it is not of great interest here since it will barely be populated during the STIRAP, and we will give a simple proof in Sec. \ref{sec:append:dark:stirap}. Similar as Eq.~(\ref{eq:dark_state:single:geo_phase}), the phase accumulated for $\ket{D_{2}}$ is calculated as
\begin{equation}\label{eq:dark_state:two:geo_phase}
\gamma_{2}^{g} = j\int\ev**{\left(\bm{e}_{\theta}\frac{\partial}{\partial\theta} + \bm{e}_{\phi}\frac{\partial}{\partial\phi}\right)}{D_{2}}\cdot \left( \bm{e}_{\theta}\partial\theta + \bm{e}_{\phi}\partial\phi \right) = j\int\ev**{\frac{\partial}{\partial\phi}}{D_{2}}\partial\phi = -\int\frac{\cos^{4}{\theta} - 2\sin^{4}{\theta}}{2 - \cos^{4}{\theta}}\partial\phi.
\end{equation}
\subsection{STIRAP and Controlled-Z gate\label{sec:append:dark:stirap}}

\begin{figure}[htbp]
    \centering
    \includegraphics{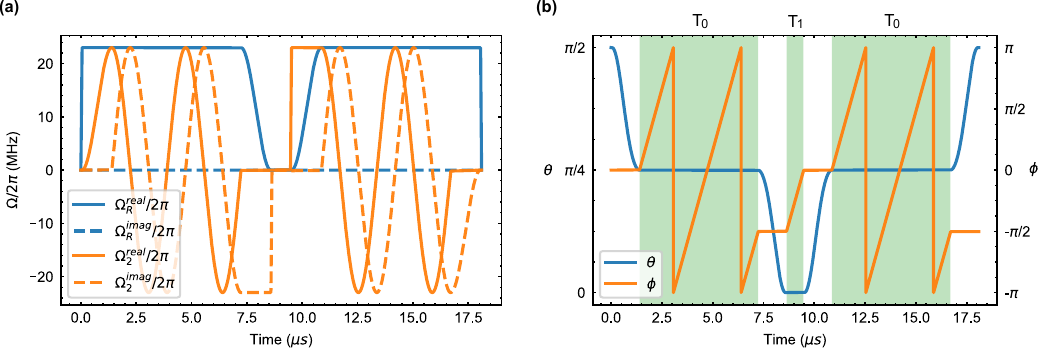}
    \caption{STIRAP pulse shapes in (a) real \& imaginary part and (b) $\theta$ \& $\phi$ representations.  Geometric phase is designed to only accumulate when $\theta$ remains constant (shaded area in (b)) in order to facilitate the calculation of overall phases in the construction of quantum gates. $T_{0}$ and $T_{1}$ above the figure indicate the duration of the shaded area underneath respectively. \label{fig:STIRAP_pulseshape}}
\end{figure}

STIRAP aims at adiabatically and coherently transferring population between quantum energy levels by applying shaped electromagnetic pulses \cite{STIRAP_PRA2007,STIRAP_PRL2008,vitanov_stimulated_2017}. Typical STIRAP pulse shapes for phase gate design used in our simulation are plotted in Fig. \ref{fig:STIRAP_pulseshape}, with both varying amplitudes and phases to adjust population transfer and phase accumulation at the same time. The rising and the falling edge of the pulse amplitude take the form of $\sin^{2}$ shape, which is adopted in \cite{STIRAP_PRA2007}. To construct the CZ gate, note that we use time-reversed symmetric pulse shapes such that in the end the population retrieves in both single- and two-spin scenarios, which corresponds to $\theta$ taking a round-trip back to its original value.

More crucial part is to quantify the amount of phases accumulated for each initial states associated with the gate construction. Here we specifically design the pulses such that the relative phase between $\Omega_{R}$ and $\Omega_{2}$ only accumulates when $\theta$ remains constant (see Fig. \ref{fig:STIRAP_pulseshape}) since this enables analytical solutions to the geometric phases defined in Eq.~(\ref{eq:dark_state:single:geo_phase}) and (\ref{eq:dark_state:two:geo_phase}). To construct the CZ gate, we have to make sure that only $\ket{11}$ undergoes a $\pi$ phase shift, which means
\begin{equation}\label{eq:dark_state:stirap:phases}
\begin{aligned}
\gamma_{1} &= \gamma_{1}^{g} = -\int\cos^{2}{\theta}\partial\phi = -\frac{1}{2}\dot{\phi}2T_{0} - \dot{\phi}T_{1} = 2k\pi,\\
\Delta\gamma &= \gamma_{2} - 2\gamma_{1} = \gamma_{2}^{g} - \phi - 2\gamma_{1}^{g} = \int\frac{2\cos^{4}{\theta}\sin^{2}{\theta}}{2 - \cos^{4}{\theta}}\partial\phi = \frac{1}{7}\dot{\phi}2T_{0} = \pi.
\end{aligned}
\end{equation}
Here we pick $k=-2$ and denote $\dot{\phi} = 2\pi \delta_{R2}$, then we obtain $T_{0} = 7 / (4\delta_{R2})$ and $T_{1} = 1 / (4\delta_{R2})$. We choose $\delta_{R2} = 0.3$ MHz for our simulation, while other parameters are summarized in the main text.

% \begin{table}[htbp]
%  \caption{\label{tab:table2}Simulation parameters (unit: GHz, except $t_{\text{rise}}$) for Fig. 4 in the main text. Note that instead of using constant values for $\Omega_{1}$ and $\Omega_{2}$, we set the detuning $\Delta=0$ and incorporate a rising time $t_{\text{rise}}$ (time spent to reach the target Rabi frequency) for the adiabatic process. We've eliminated the pure dephasing of the excited state since the STIRAP process is robust against such noises.}
%  \centering
%  \begin{ruledtabular}
%  \begin{tabular}{cccccccccc}
%    $\omega_{m}/2\pi$&$g/2\pi$&$\Delta/2\pi$&$t_{\mathrm{rise}}$&$\Gamma_{m}^{1}$&$\Gamma_{e}^{1}$&$\Gamma_{e}^{\phi}$&$\Gamma_{s}^{1}$&$\Gamma_{s}^{\phi}$\\
%    \hline
%    5.6 & 0.257 & 0 & 1.35$\mu s$ & $10^{-6}$ & 0.01 & 0 & $10^{-9}$ & $10^{-6}$\\
%  \end{tabular}
%  \end{ruledtabular}
% \end{table}

\begin{figure}[htbp]
    \centering
    \includegraphics{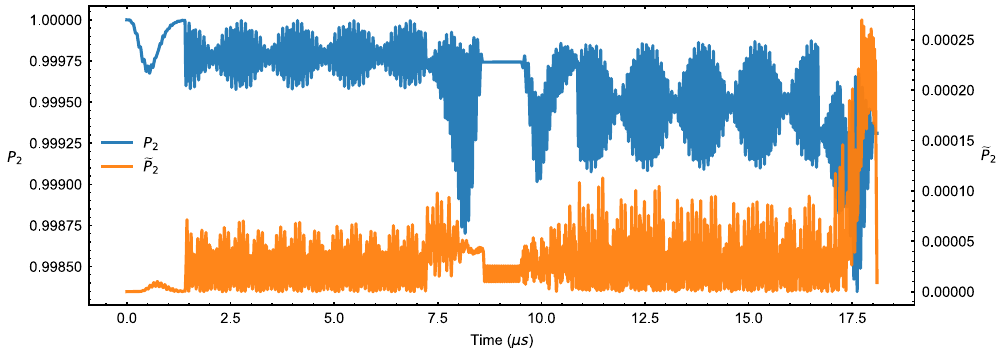}
    \caption{State leakage simulation within the dark state subspace using the pulse shape in Fig. \ref{fig:STIRAP_pulseshape} but without any decoherence. \label{fig:state_leakage}}
\end{figure}

To prove that the other dark state $\ket{\widetilde{D}_{2}}$ in Eq.~(\ref{eq:dark_state:two:D2tilde}) is barely populated, we have to investigate the dynamics in the degenerate dark subspace \cite{Wilczek_1984,STIRAP_PRA2007}. If we write our time-dependent dark state as $\ket{D(t)} = C_{2}(t)\ket{D_{2}(t)} + \widetilde{C}_{2}(t)\ket{\widetilde{D}_{2}(t)}$, we will obtain two differential equations governing the system dynamics
\begin{equation}\label{eq:dark_state:stirap:otherdarkpopeq}
\begin{aligned}
\dot{C}_{2}(t) &= -C_{2}(t)\braket{D_{2}(t)}{\dot{D}_{2}(t)} - \widetilde{C}_{2}(t)\braket{D_{2}(t)}{\dot{\widetilde{D}}_{2}(t)},\\
\dot{\widetilde{C}}_{2}(t) &= -C_{2}(t)\braket{\widetilde{D}_{2}(t)}{\dot{D}_{2}(t)} - \widetilde{C}_{2}(t)\braket{\widetilde{D}_{2}(t)}{\dot{\widetilde{D}}_{2}(t)},
\end{aligned}
\end{equation}
with initial conditions $C_{2}(0) = 1$, $\widetilde{C}_{2}(0) = 0$. Here the most important quantity is $\braket{\widetilde{D}_{2}(t)}{\dot{D}_{2}(t)}$ since it determines how fast the population gets transferred from $\ket{D_{2}}$ to $\ket{\widetilde{D}_{2}}$ if the initial state is $\ket{D_{2}}$. With Eq.~(\ref{eq:dark_state:two:D2}) and (\ref{eq:dark_state:two:D2tilde}), we can readily obtain
\begin{equation}\label{eq:dark_state:stirap:transrate}
\begin{aligned}
&\braket{\widetilde{D}_{2}}{\dot{D}_{2}} =\\
&\frac{e^{j\phi}}{\sqrt{\left( 3 + 2\cos^{2}{\theta} - 3\cos^{4}{\theta} \right)\left( 2 - \cos^{4}{\theta} \right)}} \Biggl[
\begin{aligned}[t]
&-\sqrt{2}\sin^{2}{\theta}\left( \frac{\mathrm{d}}{\mathrm{d}t}\left( \frac{\cos^{2}{\theta}}{\sqrt{2 - \cos^{4}{\theta}}} \right) + \frac{\cos^{2}{\theta}}{\sqrt{2 - \cos^{4}{\theta}}}j\dot{\phi} \right)\\
&+ \left( 2\cos^{4}{\theta} - 3\cos^{2}{\theta} \right)\left( \frac{\mathrm{d}}{\mathrm{d}t}\left( \frac{\sqrt{2}\sin^{2}{\theta}}{\sqrt{2 - \cos^{4}{\theta}}} \right) - \frac{\sqrt{2}\sin^{2}{\theta}}{\sqrt{2 - \cos^{4}{\theta}}}j\dot{\phi} \right)\\
&+ \sqrt{2}\sin{\theta}\cos{\theta}\left( 1 + \sin^{2}{\theta} \right)\frac{\mathrm{d}}{\mathrm{d}t}\left( \frac{2\sin{\theta}\cos{\theta}}{\sqrt{2 - \cos^{4}{\theta}}} \right)\Biggr].
\end{aligned}
\end{aligned}
\end{equation}
Considering the STIRAP pulse shape mentioned before, it should be verified that the contribution is small from the duration when both $\Omega_{R}$ and $\Omega_{2}$ are at their maximum ($\theta = \pi/4$) with a varying relative phase $\phi(t)$ (when geometric phase is accumulated), because for the coupling turning on/off stages, the leakage can always be mitigated by decreasing the rising time provided that the rate is bounded \cite{STIRAP_PRA2007}. To estimate the contribution of the duration when the geometric phase is accumulated, suppose the populations only manifest small perturbations around the initial conditions, which means $C_{2}(t)\sim 1$ and $\widetilde{C}_{2}(t)\sim 0$, then we can approximate the dynamics by $\dot{\widetilde{C}}_{2}(t)\sim -\braket{\widetilde{D}_{2}(t)}{\dot{D}_{2}(t)}$, with which we can get $\widetilde{C}_{2} \simeq -\eta\left( e^{j\phi} - 1 \right)$, where $\eta = \frac{2}{7}\sqrt{\frac{2}{13}}$ is computed at $\theta = \pi/4$ using Eq.~(\ref{eq:dark_state:stirap:transrate}). In this way, we obtain an upper bound on the leakage to $\ket{\widetilde{D}_{2}}$, which is $\widetilde{P}_{2} = \abs{\widetilde{C}_{2}}^{2} \simeq 4\eta^{2}\sin^{2}{\frac{\phi}{2}} \leq 4\eta^{2} = 32/637 \sim 5\%$. With numerical simulation, it is shown in Fig. \ref{fig:state_leakage} that the leakage level is far below the upper bound partly due to the periodicity originating from Eq.~(\ref{eq:dark_state:stirap:transrate}).

\subsection{Preparation of Dicke states\label{sec:append:dark:dicke}}
Here we provide more details as to the process of generating one-excitation Dicke states $D_1^N$ utilizing the STIRAP protocol, where $N$ is the spin number. Consider again the Hamiltonian
\begin{equation}\label{eq:dark_state:dicke}
\mathcal{H} = - \left( \frac{\Omega_{R}}{2}b\sum\limits_{i}\ket{e_{i}}\bra{g_{i1}} + h.c. \right) + \left( \frac{\Omega_{2}}{2}\sum\limits_{i}\ket{e_{i}}\bra{g_{i2}} + h.c. \right),
\end{equation}
where $\Omega_R = \Omega_1 g/\omega_m$. Here, we assume identical spin frequencies, coupling strengths to the phonon mode, and laser driving amplitudes. As a result, it is sufficient to work in the symmetric spin bases $\left\{ \ket{S_1}, \ket{S_2}, \ket{S_3} \right\}$, where
\begin{equation}
    \begin{aligned}
        \ket{S_1} &= \ket{1g_1\cdots g_1},\\
        \ket{S_2} &= \frac{1}{\sqrt{N}}\left( \ket{0eg_1\cdots g_1} + \ket{0g_1eg_1\cdots g_1} + \cdots + \ket{0g_1\cdots g_1e} \right),\\
        \ket{S_3} &= \frac{1}{\sqrt{N}}\left( \ket{0g_2g_1\cdots g_1} + \ket{0g_1g_2g_1\cdots g_1} + \cdots + \ket{0g_1\cdots g_1g_2} \right).
    \end{aligned}
\end{equation}
In this case, the Hamiltonian in Eq.~(\ref{eq:dark_state:dicke}) can be written in the matrix form
\begin{equation}
    \mathcal{H} = \begin{pmatrix}
        0 & -\frac{\Omega_R^*}{2}\sqrt{N} & 0\\
        -\frac{\Omega_R}{2}\sqrt{N} & 0 & \frac{\Omega_2}{2}\\
        0 & \frac{\Omega_2^*}{2} & 0
    \end{pmatrix},
\end{equation}
from which we can solve for the dark state
\begin{equation}
    \ket{D} = \left( \Omega_2\ket{S_1} + \sqrt{N}\Omega_R\ket{S_3} \right) / \sqrt{\abs{\Omega_2}^2 + N\abs{\Omega_R}^2}.
\end{equation}
Note that this is similar to the form of Eq.~(\ref{eq:dark_state:single:D1}), but with an additional $\sqrt{N}$ factor applied to the target Dicke state $\ket{S_3}$. Because of this $\sqrt{N}$ scaling factor, it sets a larger upper limit for the rising rate of $\Omega_1$ as $N$ increases while maintaining a similar adiabatic level, therefore yielding faster preparation process for the target Dicke state. This is consistent with the superradiance effect.

\section{Robustness of dark-state transfer against spectral diffusion}
Here we benchmark the robustness of a pulse shape against the spectral diffusion (SD) of the defect's optical transition by the fidelity of the single-phonon preparation process. The SD is introduced by randomizing the detunings $\Delta_{1}$ and $\Delta_{2}$ in either Eq.~(\ref{eq:odro}) or (\ref{eq:dark_state}). For simplicity, we assume that the SD is static within each realization and is balanced between left and right branch of the $\Lambda$ system. Note that other decoherence sources are not considered in this scenario.

\begin{figure}[htbp]
    \centering
    \includegraphics{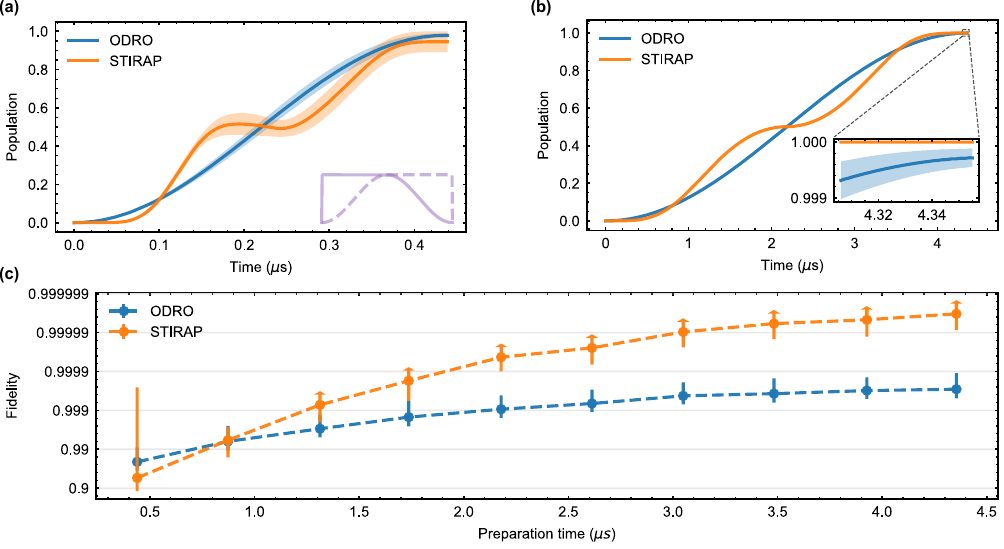}
    \caption{Benchmarking of robustness against spectral diffusion using single-phonon preparation process. We show phonon population evolution in (a) 439 $ns$ duration and (b) around 4.355 $\mu s$ duration, where the inset in (a) gives a schematic pulse shape used for the STIRAP in this simulation. For both ODRO and STIRAP, we sample the optical transition frequency from Gaussian distribution featuring a standard deviation of 20 MHz to obtain 300 different trajectories, with solid lines and shaded area representing the averaged and standard deviation of phonon population respectively. We also plot in (c) the trend of the final phonon preparation fidelity as we increase the preparation time from 439 $ns$ to 4.355 $\mu s$. Note that from the second point, the upper bound of the STIRAP fidelity is above 1 so we represent it as an arrow pointing upwards, while starting from the 4-th point (1.737 $\mu s$), STIRAP outperforms ODRO with certainty. \label{fig:SD}}
\end{figure}

As shown in Fig. \ref{fig:SD}, fidelities in both cases improve (larger averaged value and smaller standard deviation) as we increase the phonon preparation time. STIRAP manifests a more rapid improvement, almost guaranteed to be one to two orders of magnitude better than ODRO when it enters the $\mu s$ region. With short preparation times, the fidelity of STIRAP is possibly limited by lack of adiabaticity as well as not optimized pulse shapes for the rising and falling edges. In the gate design protocol of the previous section, we have a rising duration of 1.35 $\mu s$ corresponding to a single phonon preparation time of around 2.7 $\mu s$, therefore from Fig. \ref{fig:SD}(c) we demonstrate that it is justified not to include SD as a major decoherence source there.
\section{Effects of off-resonant coherent pumping}
Note that throughout the derivations above, we have neglected the off-resonant pumping of the carrier band in the $\ket{g_{i1}}\leftrightarrow \ket{e_{i}}$ transition, we here justify that this simply adds an effective frequency shift for the spin transition, which can be counteracted by adjusting the optical drive frequency detuning in the Raman process.

Analogous to Eq.~(\ref{eq:odro}), here we consider a time-dependent perturbation added to the level structure of the defects, such that
\begin{equation}\label{eq:off-res-pump}
\begin{aligned}
\mathcal{H} = {}& \mathcal{H}_{\mathrm{diag}} + \mathcal{H}_{\mathrm{offdiag}}\\
{}&\sum\limits_{i}\biggl( -\Delta_{i1}\ket{g_{i1}}\bra{g_{i1}} - \Delta_{i2}\ket{g_{i2}}\bra{g_{i2}}\biggr) + \sum\limits_{i}\left( \frac{\Omega_{i1}}{2}e^{j\omega_{m}t}\ket{e_{i}}\bra{g_{i1}} + h.c. \right).
\end{aligned}
\end{equation}
Then, according to \cite{malekakhlagh_first-principles_2020}, in order to derive an effective Hamiltonian to the second order, we need to assume a first-order generator $G_{1}(t)$ for the time-dependent SW transformation of the Floquet Hamiltonian $\mathcal{H}$ as
\begin{equation}\label{eq:td_SW_1}
G_{1}(t) = \sum\limits_{i}\left( -jx_{i}(t)\ket{e_{i}}\bra{g_{i1}} + h.c. \right).
\end{equation}
Therefore, by zeroing out the first-order effective Hamiltonian
\begin{equation}\label{eq:td_SW_2}
\mathcal{H}_{\mathrm{eff}}^{(1)} = -\dot{G}_{1} + j\left[ G_{1}, \mathcal{H}_{\mathrm{diag}} \right] + \mathcal{H}_{\mathrm{offdiag}},
\end{equation}
we obtain a solution for the coefficients $x_{i}(t) = \frac{\Omega_{i1}}{2\left( \omega_{m} + \Delta_{i1} \right)}e^{j\omega_{m}t}$, which we can use to derive the second-order effective Hamiltonian
\begin{equation}\label{eq:td_SW_3}
\begin{aligned}
\mathcal{H}_{\mathrm{eff}}^{(2)} &= -\frac{j}{2}\Bigl[ G_{1}, \dot{G}_{1} \Bigr] - \frac{1}{2}\Bigl[ G_{1}, \Bigl[ 
G_{1}, \mathcal{H}_{\mathrm{diag}} \Bigr] \Bigr] + j\Bigl[ G_{1}, \mathcal{H}_{\mathrm{offdiag}} \Bigr]\\
&= \sum\limits_{i}\left[ -\frac{j}{2}\left( x_{i}\dot{x}^{*}_{i} - \dot{x}_{i}x^{*}_{i} \right) - \Delta_{i1}\abs{x_{i}}^{2} + x_{i}\frac{\Omega_{i1}^{*}}{2}e^{-j\omega_{m}t} + x^{*}_{i}\frac{\Omega_{i1}}{2}e^{j\omega_{m}t} \right]\left( 
\ket{e_{i}}\bra{e_{i}} - \ket{g_{i1}}\bra{g_{i1}} \right)\\
&= \sum\limits_{i}\frac{\abs{\Omega_{i1}}^{2}}{4\left( 
\omega_{m} + \Delta_{i1} \right)}\left( 
\ket{e_{i}}\bra{e_{i}} - \ket{g_{i1}}\bra{g_{i1}} \right),
\end{aligned}
\end{equation}
which is essentially a frequency shift for both $\ket{g_{i1}}$ and $\ket{e_{i}}$ states. The amount $\frac{\abs{\Omega_{i1}}^{2}}{4\left( 
\omega_{m} + \Delta_{i1} \right)}$ agrees well with our simulation, and thus for the simulation of ODRO and STIRAP in the main text, we simply absorb the effect of the off-resonant coherent pumping term into the detuning coefficients $\Delta_{i1}$ and $\Delta_{i2}$.
% \section{Branching ratio?}
\clearpage
\bibliography{ref}% Produces the bibliography via BibTeX.

\end{document}